\documentclass[
 reprint,
 amsmath,amssymb,
 aps,prd,nofootinbib,showpacs
]{revtex4-1}
\usepackage{float}
\usepackage{graphicx}
\usepackage{dcolumn}
\usepackage{bm}
\usepackage{color}
\usepackage{amsmath}
\usepackage{amssymb}
\usepackage{fullpage}
\usepackage{hyperref}

\newcommand{\mpl}{m_{\mbox{\tiny{Pl}}}}
\newcommand{\Beq}{\begin{equation}\begin{aligned}}
\newcommand{\Eeq}{\end{aligned}\end{equation}}
\newcommand{\dphi}{\delta\varphi}
\newcommand{\vp}{\varphi}

\newcommand{\bk}{{\bf{k}}}
\newcommand{\bx}{{\bf{x}}}

\newcommand{\bq}{{\bf{q}}}

\newcommand{\vvp}{{\vec{\varphi}}}

\newcommand{\lsim}{\mathrel{\hbox{\rlap{\lower.55ex\hbox{$\sim$}} \kern-.3em \raise.4ex \hbox{$<$}}}}
\newcommand{\gsim}{\mathrel{\hbox{\rlap{\lower.55ex\hbox{$\sim$}} \kern-.3em \raise.4ex \hbox{$>$}}}}

\usepackage{ dsfont }

\begin{document}

\title{End of inflation, oscillons and matter-antimatter asymmetry}
\author{Kaloian D. Lozanov}\email{kd309@cam.ac.uk}
\author{Mustafa A.~Amin}\email{mustafa.a.amin@gmail.com}

\affiliation{Kavli Institute for Cosmology at Cambridge and the Institute of Astronomy, Madingley Rd, Cambridge CB3 OHA, United Kingdom}

\date{\today}
\begin{abstract}
The dynamics at the end of inflation can generate an asymmetry between particles and anti-particles of the inflaton field. This asymmetry can be transferred to baryons via decays, generating a baryon asymmetry in our Universe.  We explore this idea in detail for a complex inflaton governed by an observationally consistent --   ``flatter than quadratic"-- potential with a weakly broken global $U(1)$ symmetry. We find that most of the inflaton asymmetry is locked in non-topological soliton like configurations (oscillons) produced copiously at the end of inflation. These solitons eventually decay into baryons and generate the observed matter-antimatter asymmetry for a range of model parameters. Through a combination of three dimensional lattice simulations and a detailed linearized analysis, we show how the inflaton asymmetry depends on the fragmentation, the magnitude of the symmetry breaking term and initial conditions at the end of inflation.  We discuss the final decay into baryons, but leave a detailed analysis of the inhomogeneous annihilation, reheating and thermalization to future work.

As part of our work, we pay particular attention to generating multifield initial conditions  for the field fluctuations (including metric perturbations) at the end of inflation for lattice simulations. 

 \end{abstract}

\maketitle

\tableofcontents
\section{Introduction}
Can the observed matter-antimatter asymmetry in our Universe be connected to the dynamics at the end of inflation? An affirmative answer would allow us to connect the earliest stages of the Universe's history to this intriguing asymmetry in our Universe. 

Inflation provides the necessary initial conditions for the formation of structure in our Universe \cite{Guth:1981,Linde:1982}. However, once inflation ends the energy of the inflaton must eventually be converted into standard model particles as well as dark matter (reheating). Reheating connects inflationary physics with better understood physics of the Standard Model, possibly via intermediaries. The end of inflation and reheating can be complex, with nonlinear dynamics giving rise to a number of distinctly non-perturbative phenomena such as inflaton fragmentation, explosive particle production, defect formation etc. \cite{Kofman:1997yn, Traschen:1990sw, Amin:2011hj}. It also has a number of challenging but important observational consequences (see for example \cite{Allahverdi:2010xz}).

Experiences from our immediate surroundings, as well as from cosmological observations tell us that there are more baryons than anti-baryons in our Universe \cite{Coppi:2004za}. Moreover, the number of baryons compared to photons is extremely small \cite{Ade:2013zuv}
\Beq
\eta\approx 6\times 10^{-10}.
\Eeq 
These observations beg the question of how such an asymmetry was generated. Sakharov \cite{Sakharov:1967dj} provided the conditions necessary to generate such an asymmetry (i) departure from thermal equilibrium (ii) CP and C violation (iii) non-conservation of baryon number. Within the Standard Model, baryogenesis is difficult \cite{1985PhLB..155...36K}. A number of ideas for the generation of baryon asymmetry, with ingredients from beyond the Standard model have been put forth (see for example \cite{Dolgov:82,Abbott:1982hn,Dine:2003ax,Cline:2006ts}). Amongst the many proposals, the Affleck-Dine mechanism \cite{AffleckDine:1985} is often invoked to generate the requisite asymmetry using extra scalar fields. Such fields are easily available in supersymmetric extensions of the Standard Model. 

Recently, a variation of the Affleck-Dine mechanism using the inflaton as the Affleck-Dine scalar field was proposed in \cite{Hertzberg:2013mba, Hertzberg:2013jba} (for an earlier, related work, see \cite{Rangarajan:2001}). The authors provide an elegant analysis of asymmetry generation from the {\it homogeneous} dynamics of the inflaton at the end of inflation and provide possible particle physics embeddings to generate the observed baryon asymmetry. While the homogeneous analysis is sufficient for the quadratic inflation scenario (analyzed in detail in these papers), such an analysis is insufficient  when nonlinearities in the potential are present and lead to fragmentation of the inflaton. 

The main goal of this paper is understanding the effects of inflaton fragmentation on the generated asymmetry. We will show that when nonlinearities in the potential are present, the asymmetry can be qualitatively and quantitatively different from the homogeneous scenario. Moreover, the fragmentation leads to copious formation of pseudo-solitonic configurations (oscillons \cite{Bogolyubsky:1976yu,Gleiser:1993pt,Copeland:1995fq, Amin:2010jq,Amin:2013ika}) after inflation which lock up most of the energy density as well as the asymmetry. Some of these oscillons have an inflaton excess, others have an anti-inflaton excess. They eventually decay to generate the observed baryon asymmetry in the Universe. 

We model the inflaton as a complex scalar field with a potential motivated by monodromy inflation \cite{Silverstein:2008sg, McAllister:2008hb, Flauger:2009ab, Dong:2010in,McAllister:2014mpa}. As in \cite{Hertzberg:2013mba}, we add a small $U(1)$ breaking term to generate the inflaton/anti-inflaton asymmetry. While we focus on this particular model for concreteness, we expect our qualitative results to hold for a much broader class of ``flatter than quadratic" potentials because of the results in \cite{Amin:2011hj}.  

In this paper, we focus on the nonlinear aspects of the inflaton asymmetry generated at the end of inflation. We provide an estimate of the generated baryon-to-photon ratio under some simplifying assumptions. However, we leave the problem of detailed quark/baryon dynamics, annihilation and thermalization  for future work. The highly inhomogeneous nature of the field configurations of the inflaton, the particle physics details of decay to quarks/baryons and their diffusion along with subsequent annihilations make a more detailed calculation necessary.

We briefly review some of the previous work related to this paper. This review is not exhaustive; our aim is to try and put our work in context of previous literature. Inhomogeneous fragmentation of the inflaton and soliton formation, but without baryogenesis, has been analyzed before (for example see \cite{Enqvist:2002si, Gleiser:2010qt, Amin:2011hj}). Q-ball \cite{Coleman,Lee:1991ax} formation has also been analysed in the context of Affleck-Dine baryogenesis  with supersymmetric flat directions (for example see \cite{Enqvist:1997si, Enqvist:1997si, Enqvist:1999mv}). In our case, the inflaton acting as the Affleck-Dine field fragments into oscillons (rather than Q-balls) which carry most of the asymmetry.  The connection between Q-balls and baryogensis has been discussed extensively in the literature \cite{Enqvist:1997si}. Oscillons have been found in a number of reheating studies as well (for example \cite{Amin:2010dc,Gleiser:2010qt, Gleiser:2014ipa, Amin:2011hj, Amin:2010dc}). In this paper, our focus has been inflaton fragmentation, soliton formation and asymmetry generation within the context of the scenario in \cite{Hertzberg:2013mba}: inflationary asymmetry generation due to a small breaking of global $U(1)$ symmetry. We also believe that our paper provides the first explicit connection between oscillons and baryogenesis. For a different inhomogeneous baryogenesis scenario, see for example \cite{Cheung:2012im}. For further reviews on baryogenesis see \cite{Dine:2003ax,Mazumdar:2011zd}

The rest of the paper is organized as follows. In Sec.~\ref{sec:model} we explicitly write down the complex inflaton model along with the $U(1)$ symmetry breaking term and define the inflaton asymmetry. In Sec.~\ref{sec:dynamics} we discuss initial conditions for our lattice simulations, linear instability in the oscillating inflaton condensate and the nonlinear dynamics of the complex inflaton field. In this section we also discuss the formation of oscillons. In Sec.~\ref{sec:Asym} we show how the inflaton asymmetry is generated in the homogeneous and fully fragmented case. We discuss the dependence of the asymmetry on the parameters of the model. We also discuss the decay into baryons, as well as the relation of the baryon asymmetry to the inflaton asymmetry. We conclude in Sec.~\ref{sec:conclusions}, with a summary of our work, comments on additional observational implications and future directions. In the Appendix, we provide a formal, linearized calculation of the asymmetry.

We use $-+++$ signature for the metric, work in units where $c=\hbar=1$ and use the reduced Planck mass throughout $\mpl=1/\sqrt{8\pi G}$. We will assume an approximately Friedmann-Robertson-Walker (FRW) universe with a metric of the form\footnote{We set the two metric potentials equal to each other. This is valid for a linear calculation in both the metric and the field fluctuations for canonical scalar fields.}
\Beq
ds^2=-\left[1+2\Psi(t,\bx)\right]dt^2+a^2(t)\left[1-2\Psi(t,\bx)\right]d\bx^2,
\Eeq
where $a(t)$ is the scalefactor. We include the metric perturbations for the calculation of {\it initial conditions} for our lattice simulations. However, for subsequent nonlinear evolution after the end of inflation (on subhorizon scales), we assume an FRW metric. 


\section{Inflaton model and asymmetry}
\label{sec:model}
In this section we model the inflaton, the breaking of global $U(1)$ symmetry and define some relevant measures of the inflaton/anti-inflaton asymmetry.
\subsection{The inflaton model}
We model the inflaton as a complex scalar field $\phi$, whose action is given by
\Beq
\label{eq:action}
S&=\int\! d^4x\sqrt{-g}\left[\frac{\mpl^2}{2}R-g^{\mu\nu}\partial_\mu\phi\partial_\nu\phi^*-V(\phi,\phi^*)\right],
\Eeq
where $g_{\mu\nu}$ is the metric, $g$ is the determinant of $g_{\mu\nu}$ and $R$ is the Ricci scalar. The equation of motion of the inflaton $\phi$ is
\Beq
\label{eq:EOM}
g^{\mu\nu}\nabla_\mu\nabla_\nu\phi-\partial_{\phi^*}V(\phi,\phi^*)=0.
\Eeq
The conjugate of Eq. \eqref{eq:EOM} yields the equation of motion for $\phi^*$. 

The potential $V(\phi, \phi^*)$ consists of two parts:
\Beq
V(\phi,\phi^*)&=V_{\rm s}(|\phi|)+V_{\rm br}(\phi,\phi^*),\\
\Eeq
where $V_{\rm s}(|\phi|)$ respects the global $U(1)$ symmetry: $\phi\rightarrow e^{i\theta}\phi$. This part of the potential controls the dynamics of the field during and after inflation (though there are some corrections from $V_{\rm br}$). $V_{\rm br}(\phi,\phi^*)$ on the other hand, breaks the global $U(1)$ symmetry, and is chosen to be subdominant, at least energetically, at all times. For concreteness, we assume the following form for $V_{\rm s}(|\phi|)$: 
\Beq
V_{\rm s}(|\phi|)
&=m^2M^2\left[\sqrt{1+2\frac{|\phi|^2}{M^2}}-1\right],\\
&=\begin{cases} 
m^2|\phi|^2-\frac{m^2}{2M^2}|\phi|^4+\hdots &|\phi|\ll M \\ \\
\sqrt{2}m^2M|\phi|-m^2M^2+\hdots &   |\phi|\gg M.
 \end{cases} 
 \Eeq
During inflation $|\phi|\gtrsim \mpl\gg M\gg m$. The form of the potential is motivated by the monodromy inflation scenarios \cite{Silverstein:2008sg, McAllister:2008hb, Flauger:2009ab, Dong:2010in}. Such ``flattened" potentials are not only well motivated theoretically, but are consistent with observations \cite{Ade:2013uln}.
 
For the symmetry breaking term, $V_{\rm br}(\phi,\phi^*)$, we can choose 
\Beq
V_{\rm br}(\phi,\phi^*)&=\frac{{c}_3}{3}\frac{m^{2}}{M}\left(\phi^3+{\phi^*}^3\right).\\
\Eeq
This is the lowest dimension symmetry breaking term considered in \cite{Hertzberg:2013mba} (note that in terms of notation, our $c_3$ is different from the one defined there). The cubic power ensures that the symmetry breaking term is subdominant at late times after the end of inflation when the inflaton potential is $V_{\rm s}(|\phi|)\approx m^2|\phi|^2$. The coefficient $m^2/M$ is chosen to make $c_3$ dimensionless. For the large field values (i.e. during inflation), this symmetry breaking term might dominate unless $c_3$ is small enough. To avoid this, we must have
\Beq
c_{3}\ll \frac{1}{N}\left(\frac{M}{\mpl}\right)^{\!2},
\Eeq
where $N$ is the number of e-folds of inflation. For $N=55$ and $M=10^{-2}\mpl $ we get $c_3\ll 10^{-6}$.

However, if we do not want $c_3$ to be very small, $V_{\rm br}$ can be modified as
\Beq
\label{eq:Break}
V_{\rm br}(\phi,\phi^*)&=\frac{c_3}{3}\frac{m^2}{M}\frac{\left(\phi^3+{\phi^*}^3\right)}{f(|\phi|)},\\
f(|\phi|)&=\left(1+2\frac{|\phi|^2}{M^2}\right)^{\!2}.
\Eeq
Since $f(|\phi|)\gg 1$ during inflation, it naturally suppresses the symmetry breaking term during inflation.{\footnote{One could imagine such a factor arising due to a conformal transformation from the Jordan to Einstein frame.}} We prefer to work with this form of the symmetry breaking term since we wish to explore the $c_3$ dependence, even when $c_3$ is close to 1. While appearing innocuous here, this also leads to an extra suppression of this term in inhomogeneous regions with large field values (even after the end of inflation).

For future reference, note that at  the end of inflation $|\phi|_{\rm end}\sim \mpl\gg M$. Hence $V_s(|\phi|_{\rm end})\gg V_{\rm br}(\phi_{\rm end},\phi^*_{\rm end})$.
\subsection{Inflationary constraints}
In our potential, we have two mass scales: $m$ and $M$. If one of them is chosen, the other is determined based on the amplitude of the curvature fluctuations observed in the cosmic microwave background \cite{Ade:2013zuv}:
\Beq
\Delta_{\mathcal{R}}^2
&= \frac{1}{12\pi^2} \left(\frac{m}{\mpl}\right)^2    \left(  \frac{M}{\mpl} \right)  (2N)^{3/2},\\
&=2.2\times 10^{-9},
\Eeq 
where we have ignored the symmetry breaking term during inflation. More explicitly for $N=55$ we have
\Beq
\label{eq:mM}
m= 1.5\times 10^{-5}\left(\frac{\mpl}{M}\right)^{3/2} M.
\Eeq
For example with $M= 10^{-2}\mpl$, we get $m= 1.5\times 10^{-4}\mpl$. For the rest of the paper, we will only consider $M$ as a free parameter.

For our model (with $N=55$), we get a scalar spectral index $n_s=0.97$ and the tensor-to-scalar ratio $r_{0.002}=0.07$; consistent with Planck \cite{Ade:2013zuv} but in tension with BICEP2 \cite{Ade:2014xna}.

\subsection{Inflaton asymmetry}
\label{sec:Asym1}
The difference between the number of inflaton and anti-inflaton particles can be written in terms of the fields as follows\footnote{We define this quantity assuming an FRW metric. As we discuss later, this is justified since the asymmetry generation from fragmentation happens on subhorizon scales where the metric perturbations can be ignored.}
\Beq
\Delta N_\phi&=N_\phi-N_{\phi^*}=i\int d^3x a^3(\phi^*\dot\phi-\dot\phi^*\phi).
\Eeq
In absence of a symmetry breaking term, this number is conserved.
Using the equations of motion we get
\Beq
\frac{d}{dt}\Delta N_{\phi}=i\int d^3x a^3\left[\phi\partial_{\phi}-\phi^*\partial_{\phi^*}\right]V_{\rm br}.
\Eeq
For the particular symmetry breaking term defined in Eq.\eqref{eq:Break}, we get
\Beq
\frac{d}{dt}\Delta N_{\phi}
&=ic_3\frac{m^2}{M}\int d^3x a^3\frac{\left(\phi^3-{\phi^*}^3\right)}{f(|\phi|)},\\
&=-2c_3\frac{m^2}{M}\int d^3x a^3\frac{|\phi|^3\sin 3\theta}{f(|\phi|)},
\Eeq
where we used $\phi=|\phi| e^{i\theta}$.
We define a spatially averaged asymmetry density and a spatially averaged energy density:
\Beq
\Delta n_{\phi}(t)&\equiv \frac{{\Delta N_{\phi}}}{a^3 V_{\rm com}},\\
\bar{\rho}_\phi(t)&\equiv\frac{\int d^3x a^3\left[|\dot{\phi}|^2+a^{-2}|\nabla \phi|^2+V(\phi,\phi^*)\right]}{(a^3V_{\rm com})},
\Eeq
where $V_{\rm com}$ is the comoving volume of interest. A dimensionless ratio characterizing the inflaton asymmetry \cite{Hertzberg:2013mba} is given by
\Beq
\label{eq:Aphi0}
A_{\phi}(t)&=\frac{\Delta n_{\phi}}{(\bar{\rho}_{\phi}/m)}.
\Eeq
From now on, we will refer to this ratio as the inflaton asymmetry. For future convenience, we define another useful spatially dependent quantity, an ``asymmetry density" as follows:
\Beq
\label{eq:ADensity}
\mathcal{A}_{\phi}(t,\bx)&=\frac{\phi^*(t,\bx)\dot\phi(t,\bx)-\dot\phi^*(t,\bx)\phi(t,\bx)}{(\bar{\rho}_{\phi}/m)}.
\Eeq
\begin{figure}[t] 
   \centering
   \includegraphics[width=2in]{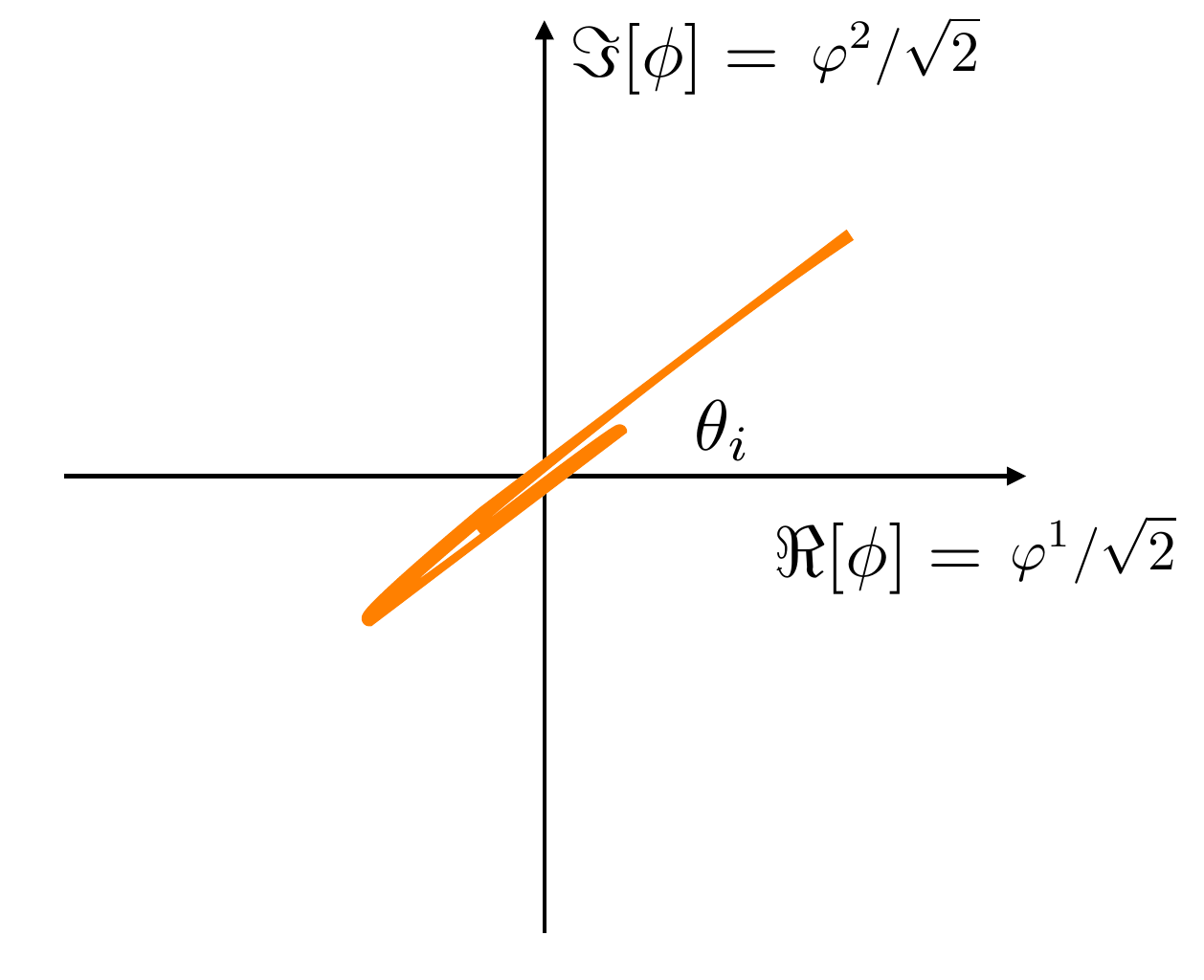} 
   \caption{A qualitative picture of the homogenous evolution of the complex inflaton field. During inflation, the symmetry breaking term is suppressed. As a result $\theta_i=\theta_{\rm inf}=$ constant. Note that the field typically starts spiraling around $|\phi|\lesssim \mpl$. }
   \label{fig:Hom}
\end{figure}

\section{Inflaton dynamics}
\label{sec:dynamics}
The equation of motion for the scalar field is given by Eq. \eqref{eq:EOM}. Both for numerical and analytic calculations, we find it is convenient to decompose the field into its ``cartesian components":
\Beq
\label{eq:CartDec}
\phi=\frac{\vp^1+i\vp^2}{\sqrt{2}}.
\Eeq
The equation of motion can then be written as 
\Beq
\label{eq:EOMCart}
g^{\mu\nu}\nabla_\mu\nabla_\nu\vp^J-\partial^J\mathcal{V}=0,
\Eeq
where $J=1,2$. Note that the covariant derivatives include the homogeneous and the inhomogeneous parts of the metric.
The potential in terms of the two fields is as follows:
\Beq
\label{eq:CartPot}
\mathcal{V}&=\mathcal{V}_{\rm s}+\mathcal{V}_{\rm br},\\
\mathcal{V}_{\rm s}&=m^2M^2\left[\sqrt{1+\frac{\delta_{IJ}\vp^{I}\vp^J}{M^2}}-1\right],\\
\mathcal{V}_{\rm br}&=\frac{c_3}{3\sqrt{2}}\frac{m^2}{M}\frac{(\vp^1)^3-3\vp^1(\vp^2)^2}{f(\vp^1,\vp^2)}.\\
\Eeq
As usual, repeated indices are summed over.

We can solve Eq. \eqref{eq:EOMCart} along with appropriate Einstein equations on a lattice, without further approximations. However, it would be a waste of computational resources to use the lattice simulations when the perturbations are small. For evolution during inflation and up to the end of inflation (or until the fluctuations in the field remain small compared to the background), we will solve the above system after linearizing in the field fluctuations. We include the metric fluctuations here since they are important for perturbations on horizon and superhorizon scales. At the end of inflation, we switch to a lattice code, which solves the full nonlinear field equation, but ignores the fluctuations of the metric. This is reasonable because although the field becomes highly nonlinear, the metric fluctuation still remains small. Moreover, we chose a simulation volume which is comparable to the comoving size of the horizon at the end of inflation since fragmentation happens on subhorizon scales (for the model considered). After the end of inflation modes never leave the horizon since the horizon grows faster than the scale factor. As a result, horizon related metric effects only matter right at the end of inflation for our simulation volume, and can be ignored thereafter. With these considerations, we include metric fluctuations in calculating the initial conditions for the fluctuations at the end of inflation, but ignore them in the lattice simulation.

\subsection{Homogeneous inflaton dynamics}
The homogeneous dynamics of the field and the metric are controlled by
\Beq
&\ddot{\vp}^I+3H\dot{\vp}^I+\partial^{I}\mathcal{V}=0,\\
&H^2=\frac{1}{3\mpl^2}\left[\frac{1}{2}\delta_{IJ}\dot{\vp}^I\dot{\vp}^J+\mathcal{V}\right].
\Eeq
Recall that $\vp^1=\sqrt{2}|\phi|\cos \theta$ and $\vp^2=\sqrt{2}|\phi|\sin \theta$.
Solving the above system numerically, we find that in the $\vp^1-\vp^2$ plane, the field maintains a constant angle during inflation when the symmetry breaking terms are subdominant: 
\Beq
\theta_{\rm inf} =\tan^{-1}(\vp^2/\vp^1)=\rm {const.}
\Eeq
After the end of inflation $\theta$ can vary, but its variation is suppressed by the size of the symmetry breaking term. In Fig. \ref{fig:Hom} we show a typical homogeneous trajectory. Note that this is a qualitative picture, the spiral is invisible for typical values of our chosen parameters.

In the usual Affleck-Dine baryogenesis, the Affleck-Dine condensate is rotating in the complex plane. In contrast, the homogeneous mode here maintains a collinear motion in the complex plane.

\subsection{Linearized perturbations}
\label{subsec:LinPert}
For this section, our results are valid for $N$ real fields. For the case at hand, $N=2$. 

When the field fluctuations are small, we can linearize the equations of motion for the field perturbations around the homogeneous values: $\vp^I+\delta\vp^I$. In Fourier space, the linearized equations of motion become
\Beq
\label{eq:linearized}
&\delta\ddot{\vp}_{\bk}^I+3H\delta\dot{\vp}_{\bk}^I+\left[\delta^I_J\frac{k^2}{a^2}+\partial^I\partial_J\mathcal{V}\right]\dphi_{\bk}^J\\
&=-2\Psi_\bk\partial^I\mathcal{V}+4\dot{\Psi}_\bk\dot{\vp}^I.
\Eeq
The potential $\Psi_\bk$ and its derivative $\dot{\Psi}_\bk$ are determined from the linearlized Einstein equations:
\Beq
&\dot{\Psi}_\bk+H\Psi_\bk=\frac{1}{2\mpl^2}\delta_{IJ}\dot{\vp}^I{\dphi}_\bk^J,\\
&\left(\dot{H}+\frac{k^2}{a^2}\right)\Psi_\bk=\frac{1}{2\mpl^2}\delta_{IJ}\left[-\dot{\vp}^I\delta\dot{\vp}_\bk^J+{\dphi}_\bk^J\ddot{\vp}^I\right].
\Eeq
One can substitute the gravitational potential $\Psi_\bk$ and its derivative $\dot{\Psi}_\bk$ into the field equations for $\dphi^J_\bk$ to get a (coupled) linear system for $\dphi^J$. Formally, we can write this linear system as
\Beq
\mathds{L}_k(t)\cdot\delta\vvp_{\bk}(t)=0,
\Eeq
where 
\Beq
\delta\vvp_{\bk}(t)&=\left[\dphi^1_{\bk}(t),\hdots,\dphi^N_{\bk}(t)\right]^T.
\Eeq
In the above equation $\mathds{L}_k(t)$ is a linear, second-order-in-time differential operator that depends on $k$ and $t$. It is a $N\times N$ matrix. For our case the operator $\mathds{L}_k$ has the form
\Beq
\mathds{L}_k\cdot\delta\vp_{\bk}(t)&=\delta\ddot{\vvp}_\bk(t)+3H\delta\dot{\vvp}_{\bk}(t)\\
&+\frac{k^2}{a^2}\delta\vec{\vp}_\bq+\mathds{M}(t)\cdot\delta\vvp_\bk\\
&+\frac{1}{\mpl^2}\left[\mathds{X}(t,k)\cdot\delta\vvp_\bk+\mathds{Y}(t,k)\cdot\delta\dot{\vvp}_\bk\right]=0.
\Eeq
The above system included scalar gravitational perturbations (terms $\propto \mpl^{-2}$). The matrices $\mathds{X}(t,k)$ and $\mathds{Y}(t,k)$ have the property $\mathds{X}(t,k),\mathds{Y}(t,k)\ll (\mpl^2H^2)(k/aH)^2$ as $k/aH\rightarrow \infty$.

The solution to this linear system can be written formally as
\Beq
\delta\vvp_{\bk}(t)
&=\sum_{n=1}^Na_{\bk n}\vec{u}_{n}(t,k)+a^*_{-\bk n}\vec{u}^*_{n}(t,k),
\Eeq
where for each $n$,
\Beq
&\vec{u}_n(t,k)=\left[u^1_n(t,k),\hdots,u^N_n(t,k)\right]^T,\\ \\
&\mathds{L}_k(t)\cdot \vec{u}_n(t,k)=0.
\Eeq
Note that the solution has $2N$ constants of integration and $2N$ ``vector" solutions. The appearance of $a^*_{-n\bk}$ is due to our assumption that $\delta\vp^J_\bk$ are Fourier transforms of real fields. 
In component form
\Beq
\label{eq:modeExp}
\delta\vp^J_{\bk}(t)
&=\sum_{n=1}^Na_{\bk n}u^J_{n}(t,k)+a^{*}_{-\bk n}u^{J*}_{n}(t,k).
\Eeq
\begin{figure}[t] 
   \centering
   \includegraphics[width=2.75in]{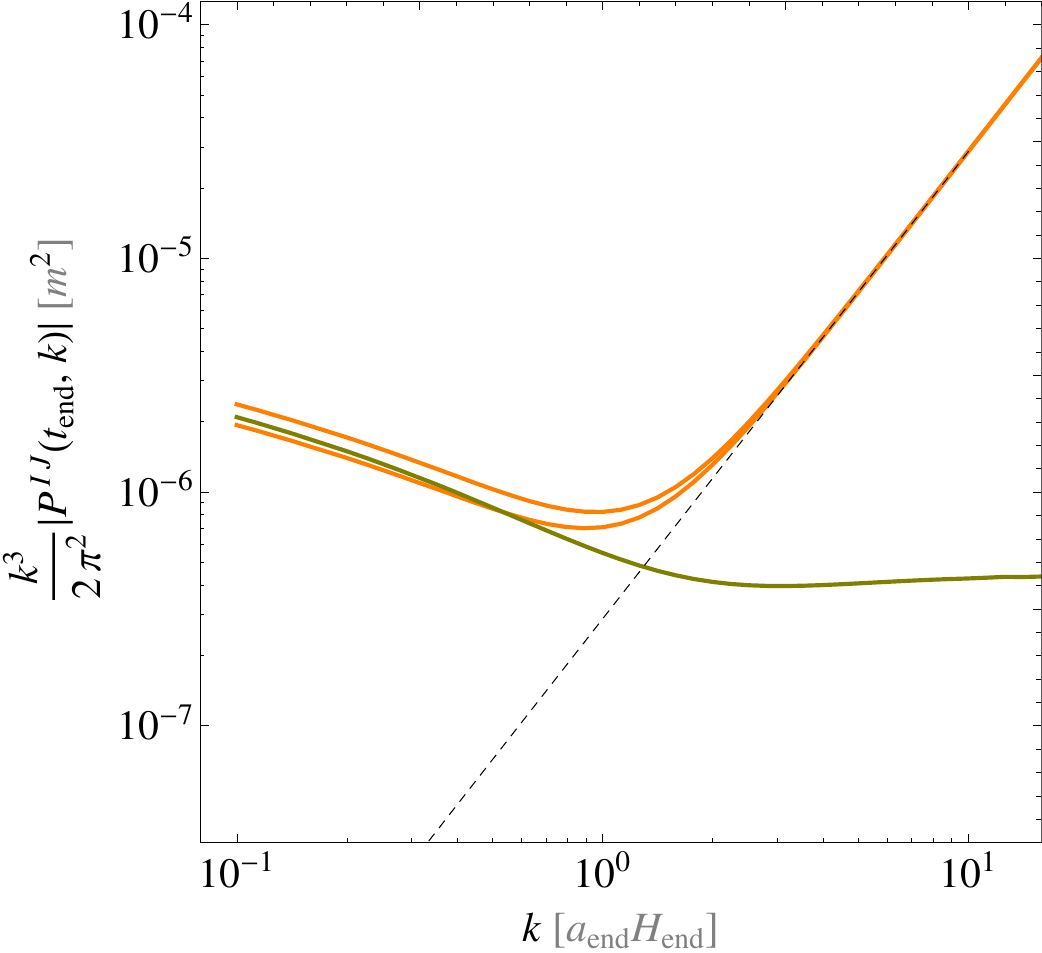} 
   \caption{Different components of the power spectra of the fields at the end of inflation (with $\theta_i= 0.7\times \pi/3 $). Inside the horizon, the diagonal components match the Minkowski space power spectrum, whereas the cross spectra are small. Outside the horizon, the perturbation spectra (diagonal spectra (orange) and cross spectra(green)) are much larger that the Minkowski space approximations (dashed line). Starting from Bunch-Davies initial conditions deep inside the horizon during inflation, we evolved the perturbations including metric perturbations self consistently. Ignoring metric perturbations underestimates the spectra on superhorizon scales.} 
   \label{fig:Spectra}
\end{figure}

\subsubsection{Quantization and power spectra}
We now follow the usual canonical quantization procedure and elevate $a_{n\bk}$ and $a^*_{n\bk}$ to operators.
\Beq
a_{\bk n}&\rightarrow \hat{a}_{\bk n},\\
a^*_{\bk n}&\rightarrow \hat{a}^{\dagger}_{\bk n},\\
\Eeq
that satisfy the following commutation relations
\Beq
\left[\hat{a}_{\bq n},\hat{a}_{\bk m}\right]&=0,\\
\left[\hat{a}_{\bq n},\hat{a}^{\dagger}_{\bk m}\right]&=\delta(\bq-\bk)\delta_{nm}.
\Eeq
Notationally, this means putting ``hats" on $\delta\vp^J_\bk$ and $\{a_{n\bk},a^*_{-n\bk}\}$ in the mode expansion in Eq. \eqref{eq:modeExp}. This expansion in terms of creation and annihilation operators is consistent with the one provided in the last chapter of \cite{2008cosm.book.....W}.\footnote{We thank D. Kaiser, M. Hertzberg and J. Karouby for discussion on two field initial conditions. A further discussion of multifield initial conditions will also be presented in an upcoming review article \cite{RevAmin:2014}.} 
Following \cite{2008cosm.book.....W}, we chose the Bunch-Davies vacuum as initial conditions. When the modes are sufficiently deep inside the horizon during inflation
\Beq
\label{eq:modeIC}
u^{J}_n(t,k)\rightarrow \delta^{J}_n\frac{\exp\left[-ik\int_{t_{\rm in}}^{t} \frac{d\tau}{a(\tau)}\right]}{(2\pi)^{3/2} a(t)\sqrt{2k}}.
\Eeq
Here, $t_{\rm in}$ stands for a time when modes of interest are deep inside the horizon. 

We can now evolve $u^J_n(t,k)$ from deep inside the horizon during inflation, through horizon crossing and up to the end of inflation. It is convenient to decompose the complex $u^J_n$ in terms of two real functions as follows.
\begin{widetext}
\Beq
\label{eq:uJnDec}
&u^J_{n}(t,k)
=\frac{1}{(2\pi)^{3/2}a(t_{\rm in})\sqrt{2k}}\left[f^J_{n}(t,k)-H(t_{\rm in})g^J_n(t,k)-i\frac{k}{a(t_{\rm in})}g^J_{n}(t,k)\right].
\Eeq
\end{widetext}
The benefit of using $f^J_n$ and $g^J_n$ is numerical ease. They are real functions satisfying
\Beq
f^J_{n}(t_{\rm in},k)&=\dot{g}^J_{n}(t_{\rm in},k)=\delta^J_{n},\\
 \dot{f}^J_{n}(t_{\rm in},k)&=g^J_{n}(t_{\rm in},k)=0.\\
\Eeq
The Bunch-Davies initial conditions are taken care of using the $k$ dependent  coefficients. Evolving $f^J_n$ and $g^J_n$ we can obtain the mode functions as well as the power spectra at any time where the linearized analysis is valid. Once we have the mode evolution, we can calculate correlation functions for the fields on any scale.

Using the commutation relations, the correlation functions for the fields are then given by\footnote{Note that our Fourier convention is $f(\bx)=\int d^3\bq f_\bq e^{i\bq\cdot\bx}$. } 
\Beq
\label{eq:2pt}
\langle0| \delta\hat{\vp}^I_\bq(t)\delta\hat{\vp}^{J\dagger}_\bk(t)|0\rangle&=\delta(\bq-\bk)P^{IJ}(t,k),
\Eeq
where
\Beq
\label{eq:PIJ}
P^{IJ}(t,k)&=\sum_{n=1}^N u_n^I(t,k) u_{n}^{J*}(t,k).
\Eeq
Note that the cross correlations are not necessarily zero and can be important, especially on superhorizon scales. This aspect has been ignored in the literature for setting up initial conditions for lattice simulations (to the best of our knowledge).\footnote{Multifield mode evolution for calculating for example, cosmic microwave background observables, has been done before. See for example \cite{Salopek:1988qh,Tent:2008, McCallister:2012,Assassi:2013gxa, Easther:2013rva}.}

For our two field model at hand, we plot the different components of the power spectra at the end of inflation in Fig. \ref{fig:Spectra}. Note that the diagonal spectra converge to the Minkowski one deep inside the horizon, whereas the cross-spectra have an interesting plateau like behavior resulting from higher order corrections in $aH/k$ (which can be derived by a careful WKB analysis):
\Beq
P^{II}(k,t)&\rightarrow \frac{1}{(2\pi)^{3} a^2(t)2k}\qquad k\gg aH,\\
P^{IJ}(k,t)&\rightarrow \mathcal{O}[(aH/k)^3]\qquad I\ne J,k\gg aH.
\Eeq
Above we assume that $k/a$ is larger than the effective mass from the potential and from gravitational effects. 
On superhorizon scales, the departure from Minkowski space power spectrum as well as the effect of metric perturbations is significant. Moreover the cross spectra are also non-negligible on superhorizon scales. 
\begin{figure*}[t] 
   \centering
   \includegraphics[width=2.5in]{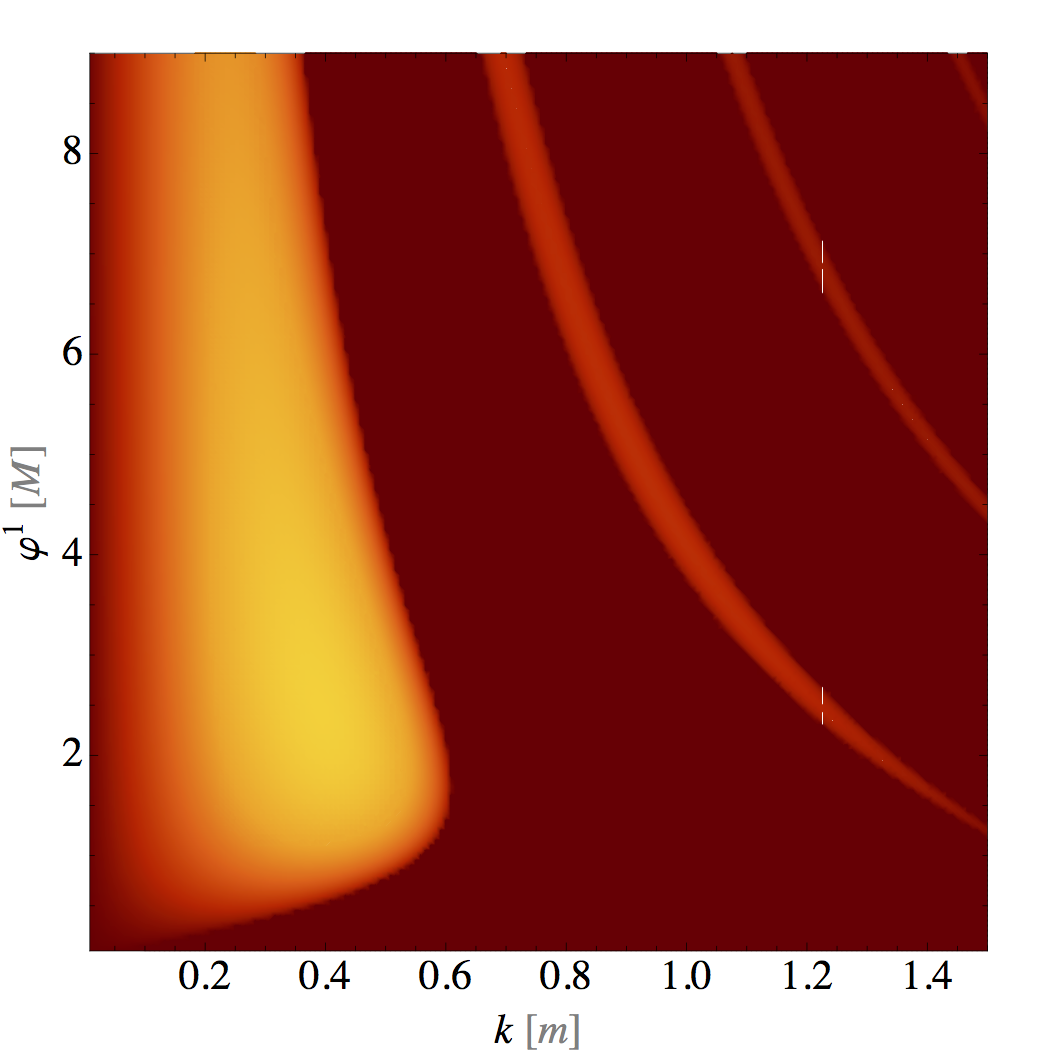} 
   \includegraphics[width=2.36in]{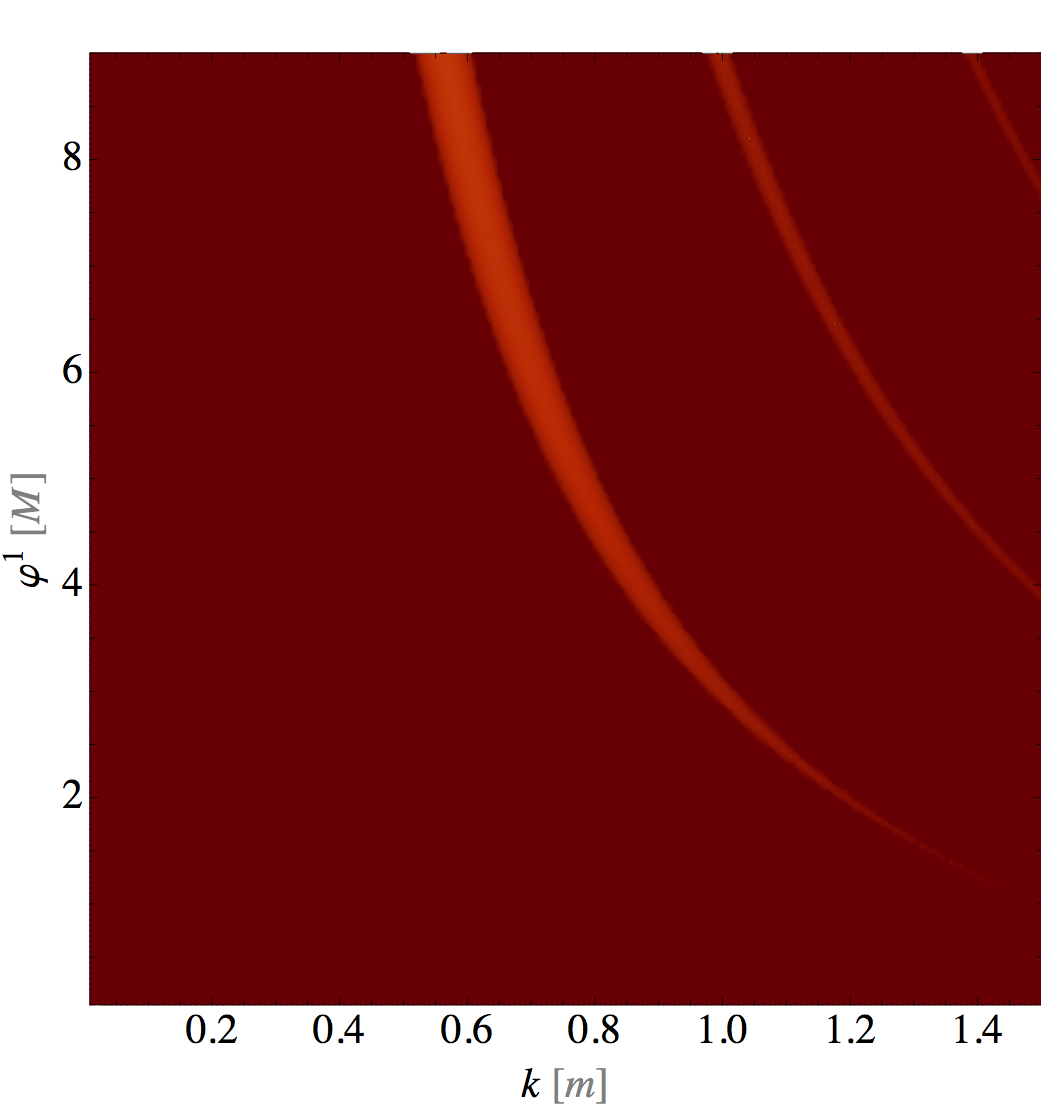} 
     \includegraphics[width=0.56in]{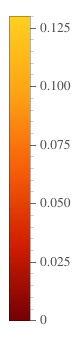} 
   \caption{Floquet charts for field fluctuations: parallel to the motion of the homogeneous field (left) and perpendicular to the motion of the homogeneous field (right). The vertical axis is the amplitude of oscillation of the homogeneous mode (assumed to be in the $\vp^1$ directions). Lighter colors correspond to unstable regions. The legend shows the magnitude of the real part of the Floquet exponent: $\Re(\mu_k)/m$. Note that the parallel perturbations have a broad, strong instability band near $k\lesssim 0.5 m$ which is not present for the perpendicular perturbations.}
   \label{fig:Floquet}
\end{figure*}

Once we obtain $P^{IJ}(t,k)$ at the end of inflation, we can populate the modes on the lattice assuming a Gaussian distribution of amplitude with uncorrelated phases. To do so, it is convenient to chose a basis where $P^{IJ}(t_{\rm end},k)$ is diagonal.   After populating the lattice using the above spectra, we rotate back to the original basis in the complex plane. This rotation back to the original basis is necessary. The final asymmetry generated depends on the breaking of $U(1)$ symmetry, and is sensitive to the angle of the homogeneous trajectory in the complex plane. 

Note that we have decided to self-consistently evolve mode functions with Bunch-Davies initial conditions from the time that modes are deep inside the horizon during inflation, up to the end of inflation. We could have chosen an instantaneous lowest energy state for each mode at the end of inflation. However, such a lowest energy state becomes ill-defined for modes outside the horizon \cite{Mukhanov:2007zz}.
While other prescriptions might be possible,  we believe that our prescription is unambiguous and physically well grounded because we start with initial conditions deep inside the horizon where all gravitational effects as well as interactions can be ignored.

Once the initial conditions are set, we use {\it LatticeEasy} \cite{Felder:2000hq} to evolve the fields. Before presenting the results of our simulations, we provide a linear analysis of the instabilities in the oscillating inflaton condensate. For the interested reader, we also provide the formalism to calculate the inflaton asymmetry based on the linearized fluctuations in the Appendix.

\subsubsection{Floquet analysis}
\label{sec:Floquet}
Soon after inflation ends, the almost homogeneous inflaton field starts oscillating around the minimum. The nonlinearities in the potential lead to an instability in the field fluctuations. The instability can be understood in terms of Floquet theory that applies to linear equations of motion with periodic coefficients. Our linearized equations of motion for the fluctuations do not have strictly periodic coefficients because of expansion as well as due to the symmetry breaking terms. For sufficiently subhorizon scales and rapid growth, we can ignore the Hubble expansion (i.e. we set $H=0$ and $a=1$ for this section). For this section we also assume that $\mathcal{V}_{\rm br}\ll \mathcal{V}_{\rm s}$. With these assumptions, as a first approximation, we arrive at
\Beq
\label{eq:FloqL}
&\delta\ddot{\vp}_{\bk}^I+\left[\delta^I_J k^2+\partial^I\partial_J\mathcal{V}_{\rm s}\right]\dphi_{\bk}^J\approx 0.
\Eeq
In absence of the symmetry breaking term, one can always rotate our field axes so that the homogeneous field is entirely along the $\vp^1$ direction. In this case the equations of motion become:
\Beq
\label{eq:FloqL12}
&\delta\ddot{\vp}_{\bk}^1+\left[k^2+\frac{m^2}{(1+(\vp^1)^2/M^2)^{3/2}}\right]\dphi_{\bk}^1\approx 0,\\
&\delta\ddot{\vp}_{\bk}^2+\left[k^2+\frac{m^2}{\sqrt{1+(\vp^1)^2/M^2}}\right]\dphi_{\bk}^2\approx 0.\\
\Eeq
As the field oscillates, the coefficients of both equations are periodic in time. According to Floquet theory, for each equation, the growing solution can be written as 
\Beq
\delta{\vp}_{\bk}^J(t,k)=\mathcal{P}^J_{1\bk}(t)e^{\mu^J_k t}+\mathcal{P}^J_{2\bk}(t)e^{-\mu^J_k t},
\Eeq
where $\mathcal{P}^J_{1,2\bk}(t)$ are periodic functions of time whereas $\mu^J_k$ are Floquet exponents. For a simple algorithm to calculate the exponents, in similar notation, see the Appendix of \cite{Amin:2011hu}. If the Floquet exponents have a real part, then we have exponentially growing solutions. We plot the real part of the Floquet exponents as a function of the amplitude of oscillations of the background field and the wavenumber $k$ in Fig. \ref{fig:Floquet}. The lighter regions are regions of instability. It is evident, that fluctuations along $\vp^1$ (i.e. parallel to the direction of the field) have broad regions of instability in contrast with the direction perpendicular to $\vp^1$.\footnote{Very recently, an analysis for linearized, uncoupled perturbations in the parallel and perpendicular directions was also provided by \cite{Hertzberg:2014jza, Hertzberg:2014iza}.} For the remainder of this section we concentrate on $\delta \vp^1_\bk$.
In an expanding FRW universe, expansion of space counteracts the exponential growth discussed above. For the instability to be efficient in an expanding universe, we have to compare the Floquet exponents (the growth rate of perturbations) to the rate of expansion. If this ratio is large compared to $1$ then we get a rapid growth of perturbations even in an expanding universe. 

For the case at hand, the nonlinearities in the potential become important when $|\phi|\sim M$. For $|\phi|=\vp^1/\sqrt{2}\sim M$, $H\sim m (M/\mpl)$. From the results of Floquet analysis we get $\Re[{\mu_{k}}]\lesssim m$ for $\vp^1\sim M$. Putting in the appropriate numerical factors, the condition for efficient growth of perturbations is given by
\Beq
\label{eq:FragEff}
\left[\frac{\Re({\mu^1_k})}{H}\right]_{\rm max}\approx \frac{1}{4}\frac{\mpl}{M}\gg 1.
\Eeq
For the factor of $1/4$ above, see \cite{Amin:2011hj}. Thus for fragmentation we need $M\ll \mpl$. From the full lattice simulations we find that fragmentation is efficient when $\mpl/M\gtrsim 40$. 

\begin{figure*}[t!] 
   \centering
   \includegraphics[width=4.5in]{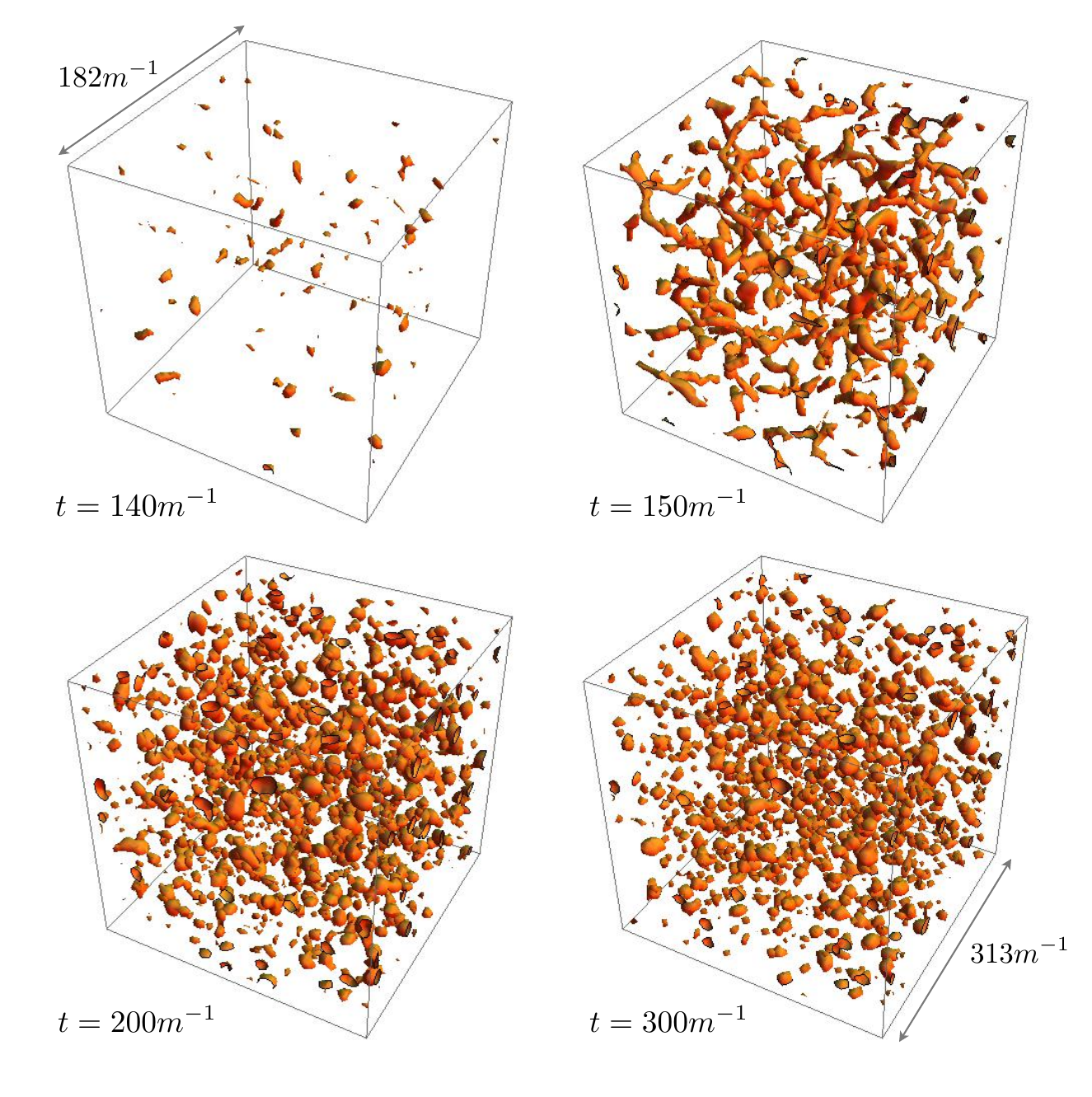} 
   \caption{The homogeneous inflaton condensate starts fragmenting within $\sim 20$ oscillations after the end of inflation. The fragmentation is driven by parametric resonance in the fluctuations along the direction of motion of the field. After the perturbations become nonlinear, localized, long-lived field configurations called oscillons form and dominate the energy density of the inflaton field. The oscillons once formed maintain a fixed size and density, and can be very long lived with lifetimes $\gg m^{-1}, H^{-1}$. They are highly over dense regions, the contours in the above plots are drawn at $5\times$ the average density. Most of the inflaton asymmetry is locked in these oscillons although they occupy a small fraction of the volume. The co-moving size of the box is comparable to the Hubble horizon at the end of inflation.}
   \label{fig:grid}
\end{figure*}
\subsection{Nonlinear dynamics}
\label{sec:NonliDyn}
\subsubsection{Inflaton fragmentation}
Soon after the beginning of parametric resonance, the perturbations become nonlinear and backreact on the homogeneous field marking a breakdown of our linearized analysis.  About 20 oscillations after the end of inflation, the inflaton field fragments. This fragmentation leads to the formation of long-lived, localized pseudo-solitonic ``lumps". The fragmentation of the inflaton and formation of these pseudo-solitonic lumps is shown in Fig.~\ref{fig:grid}.  The ``lumps'' are highly  over-dense regions, the contours in the above plots are drawn at $5\times$ the average density. Their central densities are often more than an order of magnitude above the average density. The lumps maintain a fixed central density and physical size as the universe expands.

A closer analysis of the lumps, reveals that they are ``oscillon-like" configurations \cite{Bogolyubsky:1976yu,Gleiser:1993pt,Copeland:1995fq}. Oscillons are field configurations that are localized in space and oscillatory in time. Their field configuration rather than specific parameters in the Lagrangian controls their longevity \cite{Segur:1987mg,Hertzberg:2010yz}. 

Oscillons are similar to Q-balls \cite{Coleman} in that they are localized, non-topological solitons \cite{Lee:1991ax}. However unlike Q-balls, the fields do not rotate in the complex plane and do not have a conserved charge. Below, we provide justification for why we call our ``lumps" oscillons rather than Q-balls.
\subsubsection{Oscillons vs. Q-balls}
\label{sec:OvsQ}
Ignoring the influence of the symmetry breaking terms, the general form of oscillons and Q-balls is given by

\Beq
\phi_{\rm osc}(r,t)&= \epsilon\left[f_1(r)+if_2(r)\right]\sin \omega t+\mathcal{O}[\epsilon^3], \\
\phi_{\rm Q}(r,t)&= f(r)e^{i\omega t},
\Eeq
where $\omega <m$ and $\epsilon$ characterizes the amplitude of the oscillon. Heuristically, both objects arise when the scalar potential is effectively ``shallower-than-quadratic" for some field values \cite{Lee:1991ax,Gleiser:2009ys,Amin:2013ika}. For oscillons, the field oscillates along a particular direction in the complex plane (in essentially 1 dimensional motion), whereas for Q-balls, the field rotates in the complex plane. Note that in the literature oscillons are usually defined for real fields. We have generalized this definition to a complex field. For oscillons, the higher order terms neglected here can be important when their central amplitude is large. 

To determine whether our localized overdensities in our simulation are Q-balls or oscillons, we carried out the following two tests for a sample of $10$ objects selected at random from our simulations. For the first test, we focus on the behavior of the field profiles. Note that
\Beq
|\phi_{\rm osc} (r,t)|^2&= \left[f_1^2(r)+f_2^2(r)\right]\sin^2(\omega t)+\hdots,\\
|\phi_{\rm Q}(r,t)|^2 &= f^2(r).
\Eeq 
We found that for our sample of solitons, the magnitude-squared of the field profile matched better with a sinusoidal time dependence. \footnote{We also note that the oscillons we find here have a breathing mode (as seen in \cite{Amin:2011hj}) making the higher order terms ignored above also relevant. }

Furthermore, the ratio of the real and imaginary parts of the field inside the two types of pseudo-solitons is given by
\begin{equation}
\frac{\Re(\phi)}{\Im(\phi)}\approx
\begin{cases}
{\rm const}, & \qquad \textrm{oscillons},\\
\tan (\omega t),&\qquad\textrm{Q-balls}.\\
\end{cases}
\end{equation}
Again, for our sampled objects we found that this ratio was constant, consistent with oscillons. 

For the length of the simulation, we found that our sample objects were oscillons. However, \cite{Enqvist:2002si} have argued that similar fragmentation, albeit in a different potential and without a symmetry breaking term, generates Q-balls. We cannot rule out the possibility that if one waits for a longer time ($t\gg 300 m^{-1}$) some of the oscillons will become Q-balls. 

We note that the motion of the field inside the scalar field lumps cannot be purely radial. Since in this case  the asymmetry is obviously zero. Some deviation from collinear motion in the complex plane, sourced by the symmetry breaking term and/or by nonlinear couplings between the radial and tangential directions, is necessary for there to be non-zero asymmetry. The exact nature of ``oscillon like" solutions and their corresponding asymmetry is left for future work. We will continue to call our overdensities oscillons in what follows.

Although we are dealing with a two field model (or one complex field), the dynamics is very similar to a single real field scenario discussed in \cite{Amin:2011hj}. We find that the oscillons are $\sim 10 m^{-1}$ in width with varying amplitudes $\gg M$. The fields inside oscillons oscillate in phase with a frequency $\lesssim m$. The detailed profiles of oscillons and their lifetimes \cite{Gleiser:2009ys,Salmi:2012ta,Amin:2010jq,Hertzberg:2010yz}, interactions \cite{Hindmarsh:2007jb,Amin:2013eqa}, their size distribution \cite{Amin:2010dc,Amin:2010xe} etc. will be studied elsewhere. 
\begin{figure}[t] 
   \centering
   \includegraphics[width=3in]{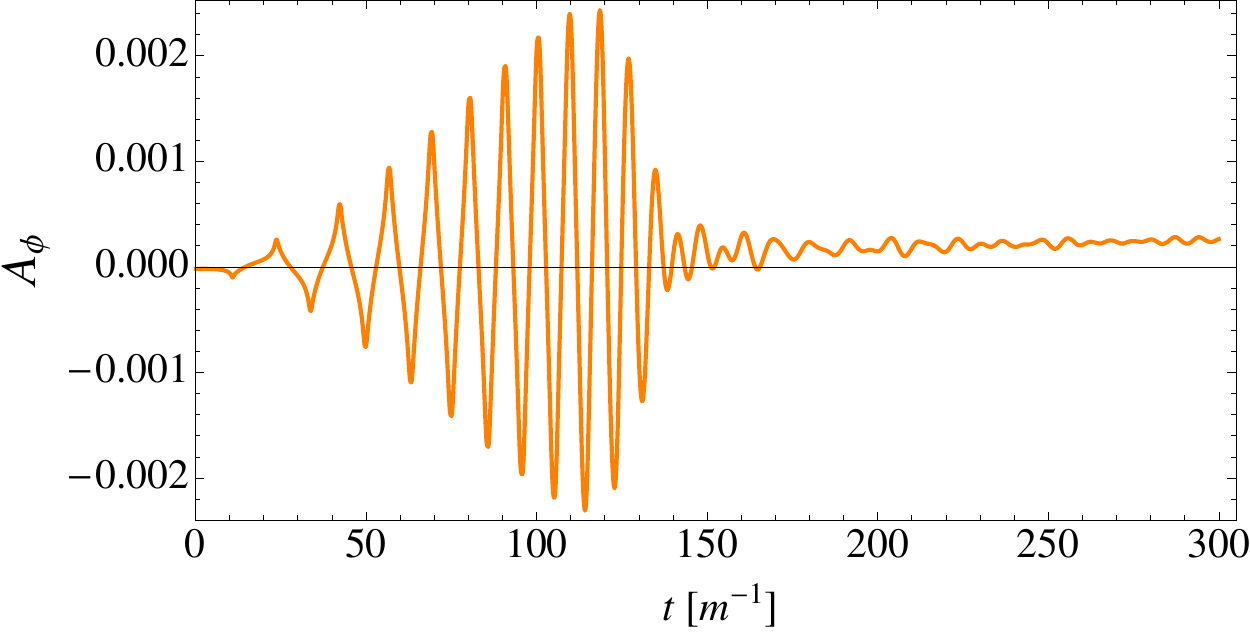} 
   \caption{Evolution of the inflaton/anti-inflaton asymmetry as a function of time. The asymmetry is zero at the end of inflation ($t=0$). Asymmetry is generated during the explosive dynamics after the end of inflation. After the inflaton fragments into localized solitons ($t\sim 150 m^{-1}$), the asymmetry settles down to a constant value. We have not checked the asymmetry for significantly longer timescales due to numerical considerations. Although not shown above, a similar plot for the asymmetry for the homogeneous case continues to show large oscillations and settles down at a much later time $t\ge 10^3 m^{-1}$.}
   \label{fig:Avst}
\end{figure}
\subsubsection{Simulation details}
We carry out a 3+1 dimensional lattice simulation of the fields in an expanding universe using a modified version of {\it LatticeEasy} \cite{Felder:2000hq}. As noted earlier, we ignore metric perturbations in the lattice code (although we include them in the initial conditions). Explicitly we solve the following equations in their discretized form
\Beq
&\ddot{\vp}^I+3H\dot{\vp}^I-\frac{\nabla^2}{a^2}\varphi^I+\partial^{I}\mathcal{V}=0,\\
&H^2=\frac{1}{3\mpl^2}\left[\frac{1}{2}\delta_{IJ}\left(\dot{\vp}^I\dot{\vp}^J+\frac{\nabla \varphi^I}{a}\cdot\frac{\nabla \varphi^J}{a}\right)+\mathcal{V}\right]_{\textrm {avg}},
\Eeq
where $I,J=1,2$ and the potential is defined in Eq. \eqref{eq:CartPot}. The right hand side of the $H^2$ equation is spatially averaged.

Our initial simulation volume was chosen to be $L=25m^{-1}$, whereas the Hubble horizon at this initial time is $H^{-1}\approx 23 m^{-1}$. We also varied the initial size of the box between $L=25m^{-1}$  and $L=50m^{-1}$ and found no significant difference between the results. This is due to the fact that resonance in our model is restricted to subhorizon scales. For $L=50m^{-1}$, the initial power spectrum on superhorizon scales is needed so as to not underestimate the power on those scales. While for this particular model, this superhorizon power does not affect the answers significantly, this need not be the case in general.

We ran our simulations for a period of 300 $m^{-1}$ after the end of inflation during which the universe expands by a factor of $\approx 12$ (and the simulation volume continues to remain sub-Hubble).  Beyond this point, we run into resolution issues, mainly because oscillons maintain a fixed physical size as the `grid' expands. It is certainly feasible to run longer, higher resolution simulations. But for our purposes, we found  a lattice with 128$^3$ points to be sufficient. We have checked that up to $t-t_{\rm end}\sim 300 m^{-1}$ there were no qualitative difference between a 256$^3$ and 128$^3$ runs. 

\begin{figure}[t!] 
   \centering
   \includegraphics[width=3in]{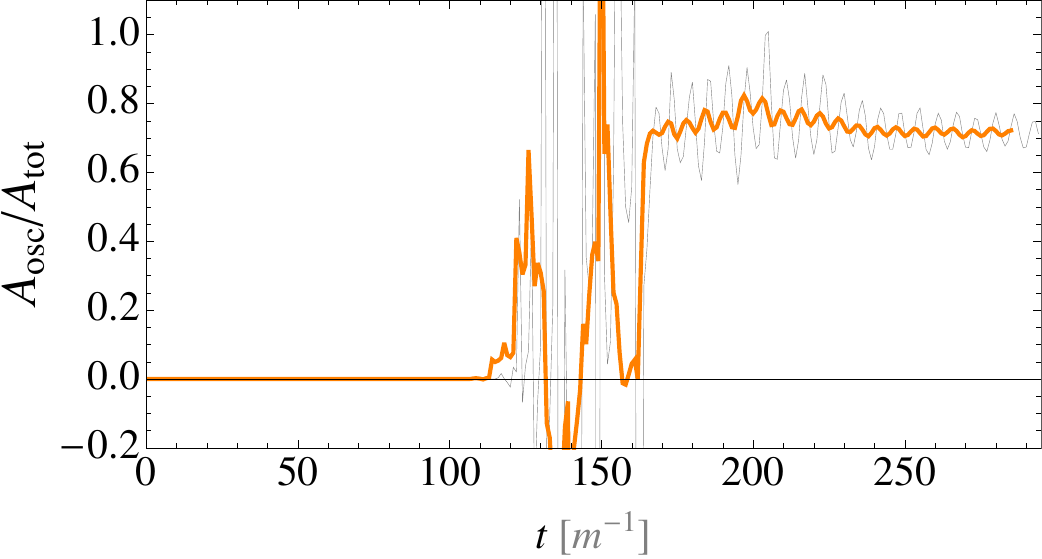} 
   \caption{The ratio of the inflaton asymmetry in regions with twice the average density to the total asymmetry (orange curve is smoothed over a few oscillations). After $t\approx 150$, the over dense regions are composed of localized pseudo-solitons (oscillons). Once oscillons are formed, most of the asymmetry is locked inside them with a final value of $A_{\rm osc}/A_{\rm tot}\approx  0.7$ . A qualitatively similar behavior is found if we consider regions with ten times the average density instead. For that case we get $A_{\rm osc}/A_{\rm tot}\approx  0.6$.}
   \label{fig:Aball}
\end{figure}

\section{Inflaton asymmetry}
\label{sec:Asym}
In the previous section we could have ignored the symmetry breaking term $\mathcal{V}_{\rm br}$ in setting up the initial conditions (though we did not). The term was included in the lattice simulations, but the results discussed so far have not depended significantly on $\mathcal{V}_{\rm br}$. In both cases the dominant contribution to the quantities of interest: the power spectrum for initial conditions and overdensities from lattice simulations, were primarily determined by the $U(1)$ symmetric piece $\mathcal{V}_{\rm s}$ of the Lagrangian. We now turn to the inflaton asymmetry, whose value is explicitly zero when evaluated using $\mathcal{V}_{\rm s}$ alone. 

As a reminder, recall that we defined a dimensionless measure of the difference between inflaton and anti-inflaton particle numbers as follows (see Sec. \ref{sec:Asym1}):
\Beq
\label{eq:asym12}
A(t)
=\frac{m}{\bar{\rho}_{\phi}(t)V_{\rm com}}\int d^3\bx \left[\dot{\vp}^{1}\vp^2-\dot{\vp}^{2}\vp^1\right].\\
\Eeq
Note that we have written the asymmetry in terms of two real fields: the real and imaginary parts of a complex inflaton.
\subsection{Asymmetry generation}
The dimensionless asymmetry (defined in Eq. \eqref{eq:asym12}) generated at the end of inflation is shown in Fig. \ref{fig:Avst}. The asymmetry is zero during inflation, and is generated after the end of inflation. After fragmentation is complete, the asymmetry settles to a fixed value. This behavior is evident in Fig. \ref{fig:Avst}. We have checked that similar qualitative behavior is seen for a wide range of parameters.

Apart from the total asymmetry, we explored the spatial distribution of the asymmetry using the asymmetry density: $\mathcal{A}(t,\bx)$ defined in equation \eqref{eq:ADensity}. After fragmentation more than $50\%$ of the asymmetry is locked in regions which are at least two times overdense compared to the average density. More precisely, for overdensities more than twice the average density, the ratio of asymmetry in the overdensity compared to the total is $70\%$, whereas for $10\times$ the average density, the ratio is $60\%$. Thus, the energy overdensity and asymmetry density are spatially localized within the oscillons. A ratio of the asymmetry inside significant overdensities (oscillons) to the total spatially averaged asymmetry is shown in Fig. \ref{fig:Aball}. Note the `jump' in overdensity and subsequent stabilization of the ratio around $t\approx 150 m^{-1}$. 
\subsection{Parameter dependence}
\label{sec:params}
In this section we describe how the asymmetry depends on the parameters in the Lagrangian based on (i)  the full lattice simulations (ii) an analysis assuming a homogeneous inflaton. For the full lattice simulations, we find that 
for $c_3\ll1$ and $M\ll \mpl$ we get
\Beq
\label{eq:Aphi}
A_{\phi}\sim \mathcal{O}[10^2]\frac{M}{\mpl}c_3^2\sin 3\theta_i,
\Eeq
where $M$ is the scale where the potential changes shape, $c_3$ is the coefficient of the symmetry breaking term, and $\theta_i$ is the initial angle of the trajectory in the $\vp^1-\vp^2$ plane. We discuss each of these dependencies in turn below. The parameter $m$ does not make an appearance because in our simulations, once $M$ is chosen, $m$ is determined based on the constraints on the amplitude of curvature fluctuations observed in the cosmic microwave background (see Eq. \eqref{eq:mM}).

For comparison with the asymmetry from lattice simulations, the homogeneous asymmetry is (for $c_3\ll 1$)
\Beq
A^{\rm hom}_{\phi}\sim \mathcal{O}[10^{-1}]\frac{\mpl}{M}c_3^2\sin 3\theta_i.
\Eeq
Note that the dependence of $\mpl/M$ is reversed between the homogeneous and the fragmented cases.
\subsubsection{Dependence on initial conditions}
\begin{figure}[H] 
   \centering
   \includegraphics[width=2.75in]{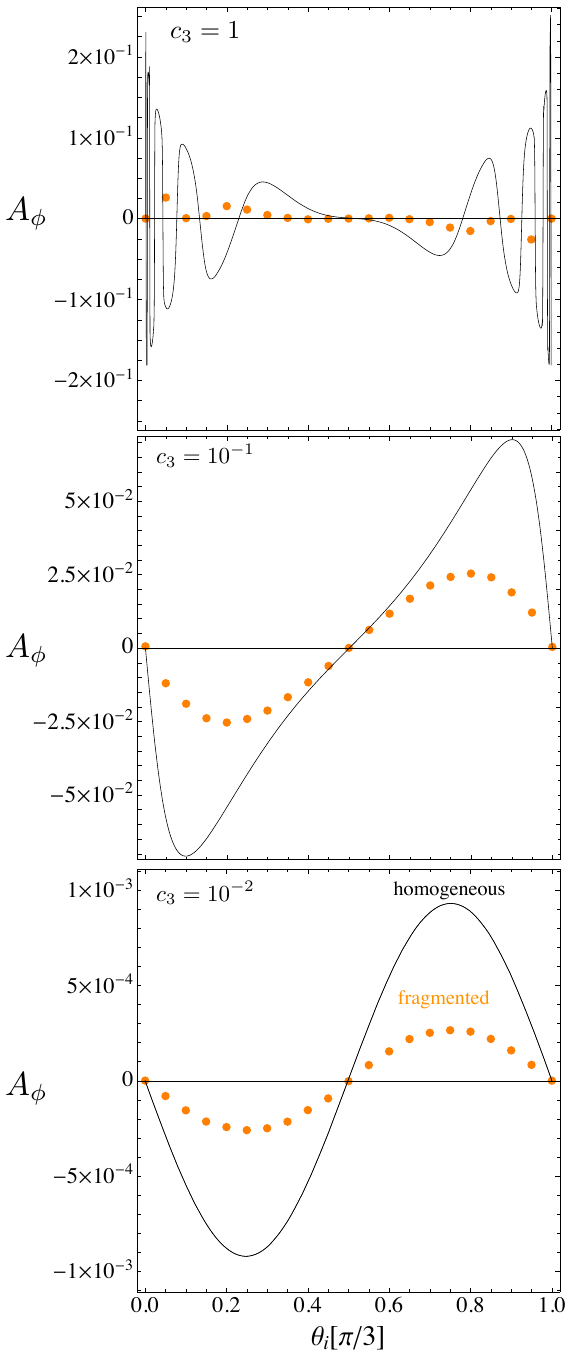} 
   \caption{Inflaton asymmetry as a function of the initial angle made by the homogeneous inflaton field in the complex plane for different values of $c_3$. The black curve corresponds to the homogeneous case, whereas the orange points are results of lattice simulations. This sinusoidal behavior seen for $c_3=10^{-2}$ is seen for all $c_3\ll1$. The $\pi/3$ period is related to the form of the symmetry breaking term. When $c_3\lesssim 1$, both the homogeneous and fragmented curves become much more complicated, no longer remaining sinusoidal. However, the $\pi/3$ period is still respected.}
   \label{fig:ATheta}
\end{figure}

The asymmetry is a strong, but simple function of the initial angle of the trajectory in $\vp^1-\vp^2$ for the homogeneous and fragmented case when $c_3\ll 1$. For small enough $c_3$, the dependence is sinusoidal $A_{\phi}\propto \sin (3\theta_i)$ where the number $3$ is related directly to the power of the fields in the asymmetry term $V_{\rm br}\propto (\phi^3+\phi^{*3})$. This dependence is shown in Fig. \ref{fig:ATheta}.  As $c_3$ approaches $1$ the behavior of the asymmetry in the homogeneous as well as the fully fragmented cases becomes much more complicated.
\subsubsection{Dependence on magnitude of the symmetry breaking term}
\begin{figure}[h] 
   \centering
   \includegraphics[width=2.7in]{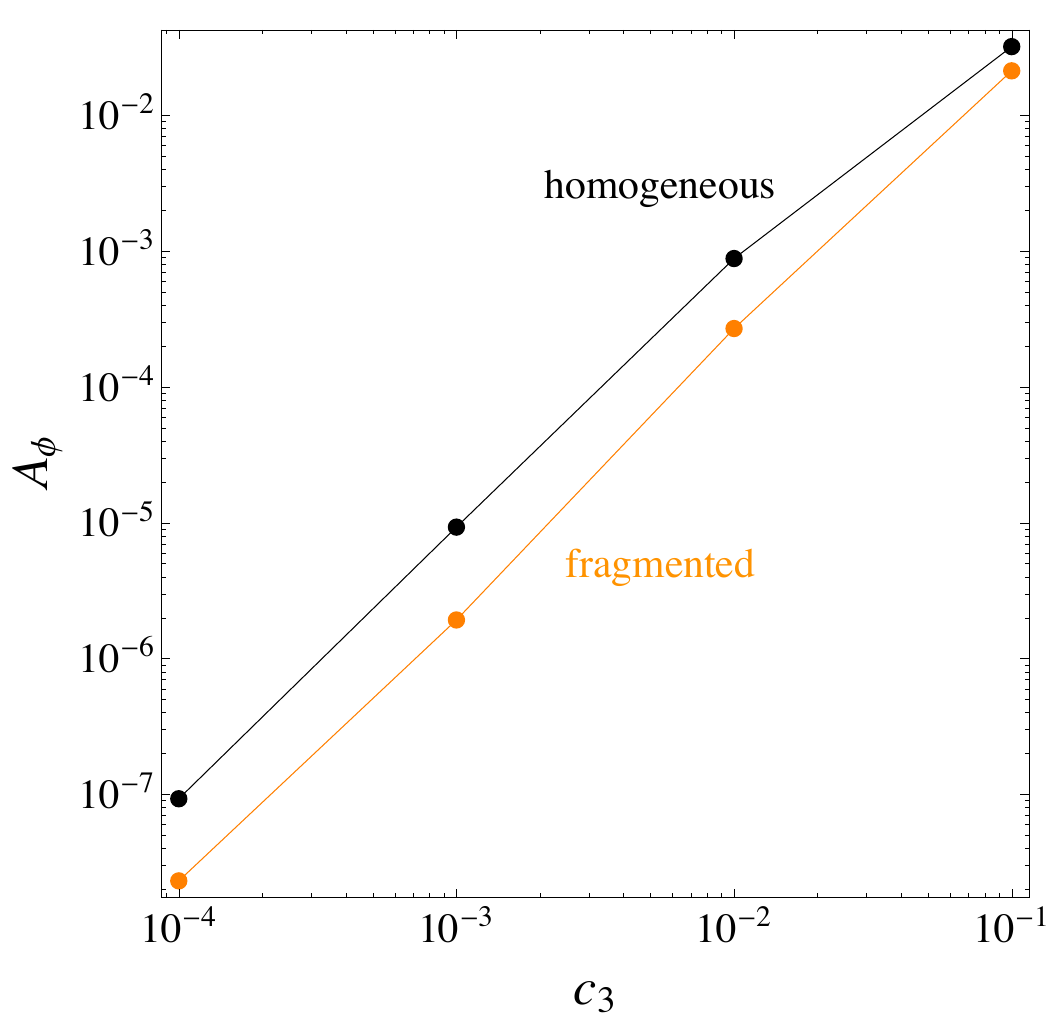} 
   \caption{Inflaton asymmetry as a function of symmetry breaking parameter, with all other parameters fixed ($\theta_{i}=0.7\times \pi/3, M=10^{-2}\mpl$). The black points correspond to the homogeneous case, whereas the orange points correspond to the results from a full lattice simulation. For $c_3\ll 1$, in both cases $A_{\phi}\propto c_3^2$, with the inhomogeneous value always being below the homogeneous one.}
   \label{fig:AvsC3}
\end{figure}

When $c_3\ll1$ we find that $A_\phi\propto c_3^2$ for both the homogeneous and fragmented cases, with a smaller value for the fragmented case. However, as $c_3$ approaches $1$, we start seeing deviations from this behavior. In Fig. \ref{fig:AvsC3} we show the dependence of $A_\phi$ on 
\begin{figure}[h] 
   \centering
   \includegraphics[width=2.75in]{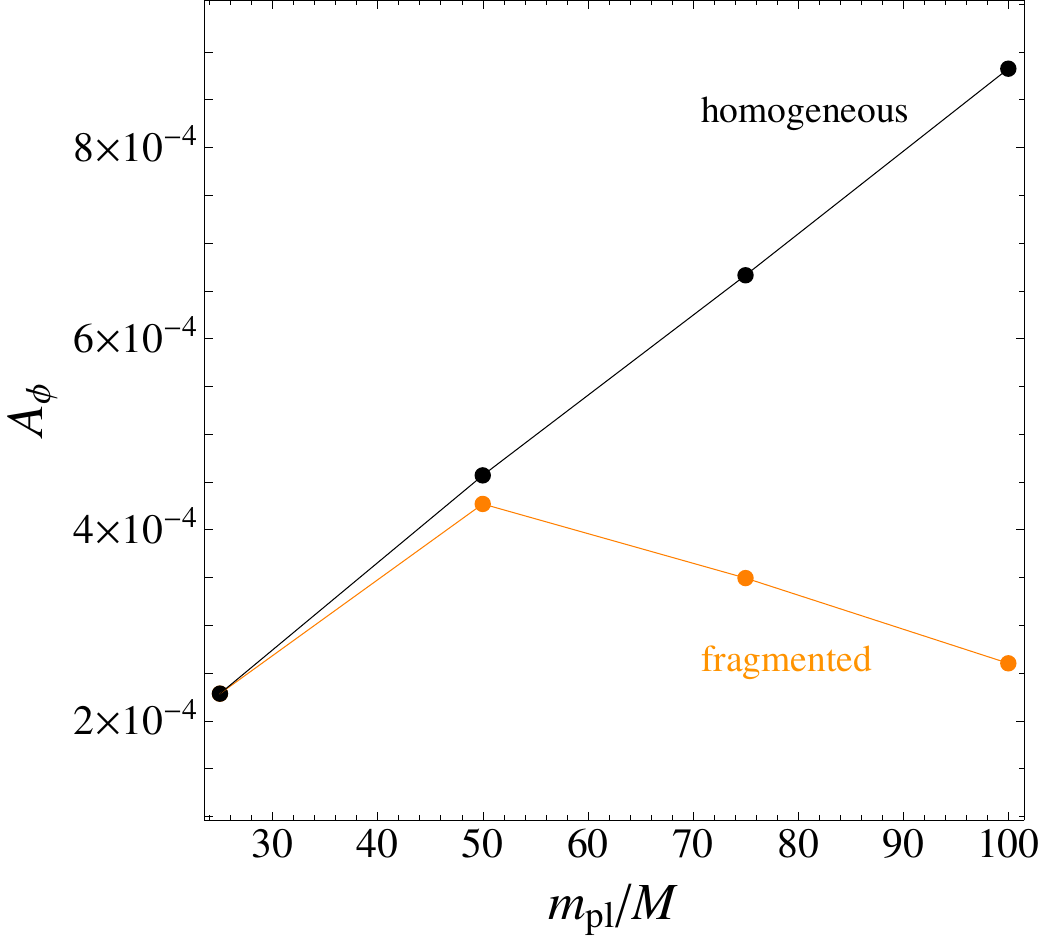} 
   \caption{Asymmetry as a function of $\mpl/M$ (with all other parameters fixed). The black points and curve correspond to the homogeneous case, whereas the orange points correspond to the results from lattice simulations. Note that the difference between the homogeneous and lattice case becomes larger and larger as $\mpl/M$ increases. The ratio $\mpl/M$ can be interpreted as the fragmentation efficiency parameter (see Eq.\eqref{eq:FragEff}). However, the symmetry breaking term also gets suppressed in the high density regions resulting from fragmentation. Hence both the fragmentation into high density regions {\it and} the suppression of asymmetry in high density regions due to the form of the symmetry breaking term determine the decrease in asymmetry as a function of $\mpl/M$ seen in the above figure.}
   \label{fig:Abeta}
\end{figure}
$c_3$ with a fixed initial angle and fixed $M$. Similar behavior is seen for different choices of $\theta_i$.

The quadratic dependence on $c_3$ helps in reducing the value of the asymmetry when 
$c_3\ll1$. This is different from the linear dependence on $c_3$ found for $m^2|\phi|^2$ inflation \cite{Hertzberg:2013mba}. The quadratic dependence on the small parameter is also different from the linear dependence discussed in \cite{Hertzberg:2014jza} where they considered a symmetry breaking term of the form $\phi^4+\phi^{*4}$. For small $c_3$, we expect symmetry should go as $\sim c_3X_1+ c_3^2X_2 +\hdots$. For the case at hand, and for the range of $c_3$ chosen, the linear term is $c_3X_1\ll c_3^2X_2$.
\subsubsection{Dependence on fragmentation}
The parameter $M$ controls the field value where the potential changes from a quadratic, to a nonlinear potential (see Sec. \ref{sec:model}). In terms of the dynamics of the inflaton the ratio $\sim \mpl/M$ controls the efficiency with which the inflaton fragments due to parametric resonance in an expanding universe (see Sec. \ref{sec:Floquet}). We plot the dependence of the asymmetry on $\mpl/M$ in Fig. \ref{fig:Abeta}. The asymmetry for the fragmented scenario starts deviating from the homogeneous case when $\mpl/M\gtrsim 50$. After that point, as the fragmentation efficiency increases the asymmetry decreases. This shows the importance of considering a full lattice simulation for calculating the asymmetry compared to the homogeneous case.

While it is clear that fragmentation plays a role, the actual reason behind the asymmetry suppression when $M\ll \mpl$ is nontrivial. The symmetry breaking term, $V_{\rm br}\propto |\phi|^3/(1+|\phi|^2/M^2)^2$. For $|\phi |\ll M$, we get $V_{\rm br}\propto M/|\phi|$. As a result in large field amplitude regions generated by fragmentation, the effect of the symmetry breaking terms (and hence the asymmetry) is suppressed. In this argument, we have assumed that the maximum  amplitude of the dense regions is independent of $M$.  
\subsection{Inflaton asymmetry to observed baryon-to-photon ratio}
\label{sec:baryons}
So far we have discussed the inflaton asymmetry in great detail. However, the observable we are ultimately interested in is the baryon asymmetry, more specifically the baryon-to-photon ratio $\eta$. The discussion below is  based on \cite{Hertzberg:2013mba}, however a fragmented inflaton introduces additional subtleties. Along with the field fragmentation, note that a matter dominated phase that results in our scenario before reheating would lead to additional nonlinear structure formation (for a low enough reheating temperature). Our main aim here is to connect the inflaton asymmetry to $\eta$ observed today.  We will comment on the differences between the homogeneous case (studied in \cite{Hertzberg:2013mba}) and our highly fragmented scenario as we present a {\it sketch} of how the decay might proceed below. 

First, at some time $t_{\phi}$ the asymmetry $A_{\phi}$ freezes out as seen in Fig. \ref{fig:Avst}. Thereafter, the inflaton/anti-inflatons decay into baryons and anti-baryons\footnote{More precisely, quarks rather than baryons since the energy scale at this time $\gg 200 \rm{MeV}$. For simplicity we will continue to refer to baryons}, by some time 
\Beq
t_{\Gamma}\sim\Gamma_{\phi}^{-1},
\Eeq
 where $\Gamma_{\phi}$ is the decay rate of the inflaton to baryons. Within any particular particle physics embedding (see for example \cite{Hertzberg:2013mba}) we can calculate $\Gamma_{\phi}$ for ``incoherent"  decay. However, the high density, coherent oscillon/Q-balls configurations might affect the decay rate \cite{Hertzberg:2010yz,Mazumdar:2002xw,Kawasaki:2012gk} significantly. We treat $\Gamma_{\phi}$  as a free parameter in what follows. \footnote{It is possible to imagine $t_{\phi}\ll t_{\Gamma}$ or they might be comparable depending on the details of the decay. }After $t_{\Gamma}$ we assume that there are no baryon number violating processes. At $t_{\Gamma}$ we have

\Beq
\label{eq:NphiToNb}
N_{b}-N_{\bar b}=b_{\phi}(N_{\phi}-N_{\phi^*})_{t_{\Gamma}},
\Eeq 
where $b_\phi=1$ or $1/3$ is the baryon number associated with the inflaton particles. For the right hand side, we assume that $(N_{\phi}-N_{\phi^*})$ is approximately constant between $t_{\phi}$ and $t_{\Gamma}$. We can  write it in terms of our asymmetry parameter as follows:
\Beq
\label{eq:NphiToAphi}
(N_{\phi}-N_{\phi^*})_{t_{\Gamma}}=\left(A_{\phi} \frac{E_{\phi}}{m}\right)_{t_{\Gamma}}=\left(A_{\phi} \frac{E_{\phi}}{m}\right)_{t_{\phi}},
\Eeq
where we used the definition of the inflaton asymmetry and $E_{\phi}$ is the energy of the inflaton field(s) in the volume of interest. The expression evaluated at $t_{\phi}$ is what we have calculated in the previous sections.

On the left-hand-side of Eq. \eqref{eq:NphiToNb}, the number of baryons minus the number of anti-baryons is fixed, after $t_{\Gamma}$. Hence this quantity is constant and can be evaluated at late times (after thermalization, and after photon number changing processes have become inefficient):
\Beq
\label{eq:NbToNg}
N_b-N_{\bar b}=(\eta N_{\gamma})_{\rm late},
\Eeq
where $N_{\gamma}$ is the number of photons for the volume of interest.  Using Eqs. (\ref{eq:NphiToNb}, \ref{eq:NbToNg} and \ref{eq:NphiToAphi}) and using spatially averaged densities, we have
\Beq
\label{eq:EtaLate1}
\eta_{\rm late}=b_{\phi}m^{-1}\frac{\left(A_{\phi}a^3\bar{\rho}_{\phi}\right)_{t_{\phi} {\textrm{  or  }} t_{\Gamma}}}{\left(a^3\bar{n}_{\gamma}\right)_{\rm late}}.
\Eeq
Note that this relates the $A_{\phi}$ we have calculated carefully to the observable $\eta_{\rm late}$. However evaluating the denominator is a bit subtle. 

After $t_{\Gamma}$ the baryons and anti-baryons produced by the decay will annihilate to produce photons. The annihilation within the solitons might be rapid, but the excess baryons/anti-baryons left over would have to diffuse through the ``plasma" to find their anti-particles to annihilate. It is unclear how long this process takes. 

Furthermore, the inflaton also decays into other particles and radiation, with the entire mix eventually reaching thermal equilibrium with a radiation like equation of state at some time $t_{\rm reh}$ (equivalently a reheating temperature $T_{\rm reh}$). It is not easy to estimate the time scale of this process and to what value $(a^3\bar{n}_{\gamma})_{\rm late}$ eventually settles.  As a result, without a careful analysis, we cannot provide a good estimate of the denominator. 

To make further progress, we have to make some strong assumptions. We will assume that 
\Beq
\label{eq:RapidTherm}
t_{\rm late}\sim t_{\rm reh}\sim t_{\Gamma}.
\Eeq
This essentially means that annihilations and thermalization happen rapidly after $t_{\Gamma}$, and photon number changing processes also become inefficient soon after.  In this case, we can evaluate the denominator and numerator at the same time $t\sim t_{\Gamma}$ and get an expression for the late time baryon-to-photon ratio $\eta_{\rm late}$ in terms of the inflaton asymmetry $A_{\phi}$, the reheating temperature $T_{\rm reh}$ and the inflaton mass $m$.

Under the assumption in Eq. \eqref{eq:RapidTherm}, we get
\Beq
\label{eq:rhoAndn}
(\rho_{\phi})_{t_{\Gamma}}\sim (\rho_{\phi})_{t_{\rm reh}}\sim & \frac{\pi^2}{30}g_* T_{\rm reh}^4,\\
(\bar{n}_{\gamma})_{\rm late}\sim(\bar{n}_{\gamma})_{t_{\rm reh}}\sim &\frac{2\zeta(3)}{\pi^2} T_{\rm reh}^3,
\Eeq
where $g_*$ is the number of relativistic degrees of freedom at that time. Using Eq. \eqref{eq:rhoAndn}  in Eq. \eqref{eq:EtaLate1} we have
\Beq
\label{eq:EtaLate2}
\eta_{\rm late}=\beta \times \frac{b_{\phi}\pi^4g_*}{60 \zeta(3)}A_{\phi}\left(\frac{T_{\rm reh}}{m}\right).
\Eeq
The factor $\beta$ is meant to account for a number of simplifications we have made such as rapid annihilation, reheating and thermalization. Assuming $g_*$ the numerical pre-factor is $\approx 160$.  The inflaton asymmetry $A_{\phi}$ is the value of the asymmetry at $t_{\phi}$ (i.e. after it settles down). The details of annihilation, decay rates etc. are hidden in $T_{\rm reh}$ and $\beta$.

We are finally ready to compute a numerical value of $\eta_{\rm late}$ and how it relates to the parameters that determine the inflaton asymmetry. Note that the observed value of $\eta_{\rm late}\approx 6\times 10^{-10}$. Hence $A_{\phi}\ll1$ and/or $T_{\rm reh}\ll m$. Both are rather natural and possible. Below we provide a combination of parameters that allows us to get the desired $\eta_{\rm late}$:
\Beq
\label{eq:etaEst}
\eta_{\rm late} \sim 10^{-9} &\beta\left(10^2\frac{M}{\mpl}\right) \left(\frac{c_3}{10^{-2}}\right)^2 \\
 &\times\left(10^{-4}\frac{\mpl}{m}\right)\left(\frac{T_{\rm reh}}{10^4 \rm{TeV}}\right)\sin 3\theta_i.
\Eeq 
In general, if we want  a high reheating temperature, $c_3$ has to be small and vice-versa.  Although we have not used that relationship above, $m$ and $M$  are related via the constraint on the amplitude of curvature fluctuations.

\section{Summary and future directions}
\label{sec:conclusions}
We have investigated the generation of matter/antimatter asymmetry from the complex and rich dynamics at the end of inflation. We have shown that in a class  of models with a complex inflaton and a small breaking of $U(1)$ symmetry, the inflaton fragments into localized soliton-like configurations called oscillons. These configurations not only contain most of the energy density of the inflaton, they also carry most of the inflaton/anti-inflaton asymmetry. The oscillons decay into baryons/anti-baryons eventually giving rise to the observed baryon-to-photon ratio in our Universe for a broad range of parameters. 

We took care in specifying multifield initial conditions on the lattice on super/subhorizon scales. Instead of using the instantaneous lowest energy state at the end of inflation, we self-consistently evolved all independent mode functions with Bunch-Davies initial conditions from deep inside the horizon during inflation up to the end of inflation. We carried out a linearized analysis of the field fluctuations with and without  gravitational perturbations where applicable (see Sec. \ref{subsec:LinPert} and the Appendix). Because of the structure of resonance in our model, the details of the initial power spectra on superhorizon scales were not relevant for us. However, we note that multifield initial conditions on horizon and superhorizon scales could be important for setting initial conditions for lattice simulations in other models. 

We carried out detailed numerical simulations using a modified version of {\it LatticeEasy} to explore the nonlinear dynamics of the inflaton field and the asymmetry generation. We explored how the asymmetry depends on the parameters of the Lagrangian, as well as the fragmentation dynamics. We found that the fragmentation does affect the inflaton asymmetry significantly. The value of the asymmetry as well as its spatial distribution are qualitatively and quantitatively different from the homogeneous case. In general, the asymmetry in the fragmented case is smaller than the one derived by ignoring the fragmentation. Inspite of the complex dynamics, we were able to provide a simple (empirical) formula  for the inflaton asymmetry, expressing it in terms of the parameters of the Lagrangian and initial conditions in a physically transparent manner (see Eq. \eqref{eq:Aphi}). 

While we provided a detailed analysis of the asymmetry generation in the inflaton, we provided a comparatively simple analysis of the decay to quarks/baryons. How this decay takes place in a highly inhomogeneous inflaton field configuration, and the details of subsequent annihilation of the quarks/anti-quarks (baryons/anti-baryons) is left for future work. We provided an estimate for the baryon-to-photon ratio (see Eqns. \eqref{eq:EtaLate2} and \eqref{eq:etaEst}) under simplified assumption of rapid thermalization (amongst others). This estimate should be checked by a detailed analysis of the inflaton decay, inhomogeneous annihilation and subsequent thermalization.

On the theoretical side a few additional problems need to be addressed. While we argued heuristically for the form of the inflaton asymmetry, a more detailed understanding is needed. We have not explored the properties of oscillons generated here in detail. Their lifetimes, distribution of amplitudes, sizes and interactions would be useful. Importantly, longer time-scale simulations (with an initial higher resolution) are needed to quantify the long term behavior of the asymmetry.  It would be a useful check to carry out these simulations using other existing codes (besides {\it LatticeEasy}), each with their own benefits  \cite{Frolov:2008hy,Easther:2010qz, 2011PhRvD..83l3509H,2010CoPhC.181..906S,2012JCAP...04..038S}.

\subsection{Additional observational consequences}
Beyond the baryon-to-photon ratio, the scenario for baryogenesis is rich in terms of other potential observational implications.  We briefly discuss a few of them below.

Isocurvature modes are generated during inflation due to the presence of the light ``angular" component of the complex field \cite{Hertzberg:2013mba}. For our model, this leads to an isocurvature fraction,
$\alpha_{II}\sim 2.6\times 10^{-4}$, which is two orders of magnitude below the current constraints \cite{Ade:2013uln}. Note that these isocurvature modes are not due to fragmentation.\footnote{Note that in a number of Affleck-Dine Baryogeneis scenarios the isocurvature modes are unacceptabely large for high energy scale inflation (see for example \cite{Harigaya:2014tla}). However, the large vacuum expectation value of the inflaton field (which doubles as the Affleck-Dine field) suppresses the isocurvature modes \cite{Hertzberg:2013mba}.}. Another connection between baryogenesis {\it during} inflation and isocurvature modes is discussed in  \cite{BasteroGil:2011cx,Bastero-Gil:2014oga}.

The initial fragmentation, and soliton formation can lead to the generation of gravitational waves (see for example \cite{Chiba:2009zu,Kusenko:2008zm, Zhou:2013tsa}).  In addition, a long phase of soliton domination leads to a matter dominated expansion history before reheating takes place. This change in the expansion history affects the mapping of modes between horizon exit during inflation and  re-entry at late times, thus affecting our interpretation of inflationary observables \cite{Adshead:2010mc,Martin:2010kz, Easther:2013nga}. The long matter dominated phase also leads to additional gravitational structure formation in the early Universe before the inflaton decays away.

The solitons found in the simulation might be extremely long lived, serving as dark matter candidates \cite{Kusenko:2004yw}\footnote{These authors considered Q-balls rather than oscillons. Q-balls are likely to live longer than oscillons because of their (approximately) conserved $U(1)$ charge.} or they might decay into dark matter \cite{Shoemaker:2009kg}. 

The inhomogeneous annihilation, if it is inefficient might  lead to signatures during big bang nucleosynthesis or in the late Universe \cite{Khlopov:2004rw}. We hope that our work will motivate a more detailed analysis of inhomogeneous decay, annihilation and subsequent thermalization in similar models.

\begin{acknowledgments}
We would like to thank J. Pritchard,  G. Efstathiou, M. Hertzberg, D. Kaiser, J. Karouby and P. Adshead for  useful conversations. MA acknowledges the support of a Kavli Fellowship.
\end{acknowledgments}

\appendix
\section{``Linearized''  asymmetry calculation}
We can use the linearized equations of motion for $u^{J}_n(t,k)$ to calculate the inflaton asymmetry up to the point where the nonlinearities become important. Recall that
\Beq
A_\phi(t)
&=i\frac{m}{\bar{\rho}_{\phi}(t)a^3(t)V_{\rm com}}\int d^3\bx a^3\left[\phi^*\dot{\phi}-\phi\dot{\phi}^*\right],\\
\Eeq
We can also write these expressions in terms of the real and imaginary parts of the field (see Eq. \eqref{eq:CartDec}):
\Beq
A_\phi(t)
&=\frac{m}{a^3(t)\bar{\rho}_{\phi}(t)V_{\rm com}}\int d^3\bx a^3\left[\dot{\vp}^{1}\vp^2-\dot{\vp}^{2}\vp^1\right].\\
\Eeq
Dividing the field into a homogeneous background and perturbations $\vp^{J}(t,\bx)=\bar{\vp}^J(t)+\dphi^J(t,\bx)$, the asymmetry can then be written as
\Beq
A_\phi(t)
&=\bar{A}_\phi(t)+{\delta A}_\phi(t),
\Eeq
where
\begin{widetext}
\Beq
\bar{A}_\phi(t)&=\frac{m}{\bar{\rho}_\phi(t)}\left[\dot{\bar{\vp}}^{1}(t)\bar{\vp}^2(t)-\dot{\bar{\vp}}^{2}(t)\bar{\vp}^1(t)\right],\\
\delta A_\phi(t)
&=\frac{m}{\bar{\rho}_{\phi}(t)V_{\rm com}}\int d^3\bx \left[\delta\dot{\vp}^{1}(\bx,t)\delta\vp^2(\bx,t)-\delta\dot{\vp}^{2}(\bx,t)\delta\vp^1(\bx,t)\right],\\
&=\frac{m}{2\bar{\rho}_{\phi}(t)V_{\rm com}}\int d^3\bx {d^3\bk}{d^3\bq}e^{-i(\bk-\bq)\cdot \bx}\left[\delta\dot{\vp}_{\bk}^{1*}(t)\delta\vp^2_\bq(t)-\delta\dot{\vp}_{\bk}^{2*}(t)\delta\vp^1_\bq(t)+\rm{c.c}\right].
\Eeq
\end{widetext}
In the second line, there are no terms linear in $\delta\vp$, since we assume that $\int d^3\bx \delta \vp =0$.
In the last step by adding the complex conjugate we make the ``reality" of $A_\phi(t)$, which follows from the reality of $\vp^J$, manifest. 

Using the correlator in Eq. \eqref{eq:2pt}, as well as
 \Beq
\langle\delta \dot{\hat{\vp}}^I_\bk(t) \delta\hat{\vp}^{J\dagger}_\bq(t)\rangle&=\delta(\bk-\bq)\sum_{n=1}^2\dot{u}^I_{n}(t,k)u^{J*}_{n}(t,k).
\Eeq
the expectation value of the asymmetry operator $\delta \hat{A}_\phi$ is given by
\begin{widetext}
\Beq
\langle{\delta \hat{A}}_\phi(t)\rangle
&=(2\pi)^3\frac{m}{\bar{\rho}_{\phi}(t)}\sum_{n=1}^2\int d\ln k\frac{k^3}{2\pi^2} \left[\dot{u}_{n}^{1}(t,k)u^{2*}_{n}(t,k)-\dot{u}_{n}^{2}(t,k)u^{1*}_{n}(t,k)+\rm{c.c}\right],\\
&=\frac{m}{2\bar{\rho}_{\phi}(t)}\sum_{n=1}^2\int \frac{d\ln k}{2\pi^2}\frac{k^2}{a(t_{\rm in})^2} \left[\{\dot{f}_{n}^{1}(t,k)-H(t_{\rm in})\dot{g}^1_n(t,k)\}\{f^{2}_{n}(t,k)-H(t_{\rm in})g^2_n(t,k)\}\right.\\
&\quad\left.+\frac{k^2}{a^2(t_{\rm in})}\dot{g}_{n}^{1}(t,k)g^{2}_{n}(t,k)-1\leftrightarrow2\right],\\
\Eeq
\end{widetext}
where in the second line we used our decomposition of $u^J_n$ described in Eq. \eqref{eq:uJnDec}.
 This expression is now well suited for calculating the asymmetry parameter using the linearized equations of motion.


\end{document}